\documentclass[acmsmall]{acmart}
\usepackage{subcaption}
\usepackage{enumitem}
\usepackage{multirow}
\AtBeginDocument{%
  }
\setcopyright{acmcopyright}
\acmJournal{TOIS}
\copyrightyear{2018}
\acmYear{2018}
\acmDOI{XXXXXXX.XXXXXXX}

\acmConference[Conference acronym 'XX]{Make sure to enter the correct
  conference title from your rights confirmation emai}{June 03--05,
  2018}{Woodstock, NY}
\acmPrice{15.00}
\acmISBN{978-1-4503-XXXX-X/18/06}

\begin{document}

\title{Counterfactual Explanation for Fairness in Recommendation}


\author{Xiangmeng Wang}
\authornote{Equal contribution.}
\email{xiangmeng.wang@student.uts.edu.au}
\orcid{0000-0003-3643-3353}
\affiliation{
\institution{Data Science and Machine Intelligence Lab, University of Technology Sydney}
  \city{Sydney}
    \country{Australia}
}

\author{Qian Li}
\authornotemark[1]
\authornote{Corresponding author: qli@curtin.edu.au}
\email{qli@curtin.edu.au}
\orcid{0000-0002-8308-9551}
\affiliation{
  \institution{School of Electrical Engineering Computing and Mathematical Sciences, Curtin University}
  \city{Perth}
  \country{Australia}}
  
\author{Dianer Yu}
\email{Dianer.Yu-1@student.uts.edu.au}
\orcid{0000-0001-6376-9667}
\affiliation{
  \institution{Data Science and Machine Intelligence Lab, University of Technology Sydney}
  \city{Sydney}
    \country{Australia}
}

\author{Qing Li}
\email{qing-prof.li@polyu.edu.hk}
\orcid{0000-0003-3370-471X}
\affiliation{
  \institution{Hong Kong Polytechnic University}
  \country{Hong Kong}
}

\author{Guandong Xu}
\authornote{Corresponding author: guandong.xu@uts.edu.au}
\email{Guandong.Xu@uts.edu.au}
\orcid{0000-0003-4493-6663}
\affiliation{
  \institution{Data Science and Machine Intelligence Lab, University of Technology Sydney}
  \city{Sydney}
    \country{Australia}
}

\renewcommand{\shortauthors}{X. Wang and Q. Li, et al.}

\begin{abstract}
Fairness-aware recommendation eliminates discrimination issues to build trustworthy recommendation systems.
Explaining the causes of unfair recommendations is critical, as it promotes fairness diagnostics, and thus secures users' trust in recommendation models.
Existing fairness explanation methods suffer high computation burdens due to the large-scale search space and the greedy nature of the explanation search process. 
Besides, they perform score-based optimizations with continuous values, which are not applicable to discrete attributes such as gender and race.
In this work, we adopt the novel paradigm of counterfactual explanation from causal inference to explore how minimal alterations in explanations change model fairness, to abandon the greedy search for explanations. 
We use real-world attributes from Heterogeneous Information Networks (HINs) to empower counterfactual reasoning on discrete attributes.
We propose a novel \emph{Counterfactual Explanation for Fairness (CFairER)} that generates attribute-level counterfactual explanations from HINs for recommendation fairness.
Our \emph{CFairER} conducts off-policy reinforcement learning to seek high-quality counterfactual explanations, with an attentive action pruning reducing the search space of candidate counterfactuals.
The counterfactual explanations help to provide rational and proximate explanations for model fairness, while the attentive action pruning narrows the search space of attributes.
Extensive experiments demonstrate our proposed model can generate faithful explanations while maintaining favorable recommendation performance. 
We release our code at \url{https://anonymous.4open.science/r/CFairER-anony/}.
\end{abstract}



\received{20 February 2007}
\received[revised]{12 March 2009}
\received[accepted]{5 June 2009}

\maketitle

\section{Introduction}

Recommendation system (RS) as an information filtering tool has been a core in online services, e.g., e-commerce~\cite{covington2016deep,he2021click}.
It helps users discover their preferred items and benefit content providers to profit from item exposures.
Despite the huge benefits, fairness issues refer to unfair allocations (i.e., exposures) of recommended items~\cite{li2022fairness}, caused by, e.g., gender discrimination, have attracted increasing attention in RS.
Fairness-aware recommendation~\cite{fu2020fairness} has emerged as a promising solution to prevent unintended discrimination and unfairness in RS.
It aims to find feasible algorithmic approaches that reduce the fairness disparity of recommendation results.
Explaining why fairness disparity appears, i.e., \emph{what causes unfair recommendation results}, would enhance the design of fairness-aware recommendation approaches by promoting model transparency and tracking unfair factors.

There are a few fairness explanation studies in the literature, which are mainly categorized as feature-based and aspect-based methods.
Feature-based methods estimate the contribution scores of numerical features that impact model fairness.
For instance, Begley et al.~\cite{begley2020explainability} explore fairness explanations based on Shapley value estimation for the classification task. 
They calculate Shapley values of every input features to reflect their significance and then generate explanations based on calculated values.
However, this method is not applicable for deep recommendation models (e.g., neural networks~\cite{chen2018neural,he2017neural}), as the high complexity of Shapley value estimation becomes the major burden when input features are in high dimension and sparse. 
Another branch of aspect-based methods mainly perturbs user/item aspect scores and optimizes an explanation model to find perturbed aspects that affect the model fairness as explanations.  
For example, Ge et al.~\cite{DBLP:conf/sigir/GeTZXL0FGLZ22} perturb aspect scores within pre-defined user-aspect and item-aspect matrices and feed the perturbed matrices into a recommendation model.
Those perturbed aspects that alter the fairness disparity of the recommendation model are considered aspect-based explanations.
However, the perturbation space grows exponentially as the number of aspects increases, resulting in a large-scale search space to seek explanations.

The above fairness explanation methods suffer below issues: 
1) These feature/aspect-based methods usually incur high computational costs due to the high dimensionality of search space and ultimately result in sub-optimal explanations. 
Besides, these methods are presented with the greedy nature of the explanation search process. 
They optimize explanation models using greedy feature/aspect scores as significance criteria and select top features/aspects as explanations, which might have the risk of introducing pseudo-explanations.
2) These score-based optimizations can only deal with continuous attributes and thus are not well-suited for handling discrete attributes.
For example, assigning a continuous value, such as \emph{gender}=0.19, to the discrete \emph{gender} attribute is impractical in constructing explanations and provides no valuable clue to improve the explanation.
Worse still, discrete attributes are frequently used in real-world recommendation models, as user and item profiles for training models are often generated through data tagging~\cite{guan2010document} on discrete attributes. 
For instance, movie recommendations~\cite{reddy2019content,goyani2020review,zhao2016matrix} usually rely on movies tagged with discrete attributes such as genre, language, and release location. 
Consequently, score-based optimizations have limited capability in handling discrete attributes that are frequently encountered in recommendation scenarios.

Unlike previous works, we resort to counterfactual explanations~\cite{verma2020counterfactual} derived from causal inference to tackle the above issues.
Counterfactual explanations address the fundamental question: \emph{what the model fairness would be if a minimal set of factors (e.g., user/item features) had been different}~\cite{verma2020counterfactual}.
In other words, they provide ``what-if'' explanations to determine the most vital and essential (i.e., \emph{minimal}) factors that change model fairness. 
Unlike existing feature/aspect-based methods with greedy explanations, counterfactual explanations have the advantage of always being minimal w.r.t. the generated explanations and are faithful to model fairness changes.
Moreover, we leverage real-world attributes from Heterogeneous Information Networks (HINs)~\cite{wang2020joint}, for counterfactual reasoning when dealing with discrete attributes.
In contrast to value-based features and aspects, real-world attributes residing in HINs are presented as discrete nodes, with edges representing their connections.
By utilizing attributes from HINs, we can overcome the limitation of score-based optimizations to directly measure whether the removal of specific attributes changes the model's fairness.

\begin{figure}[h]
    \centering
    \includegraphics[width=0.9\textwidth]{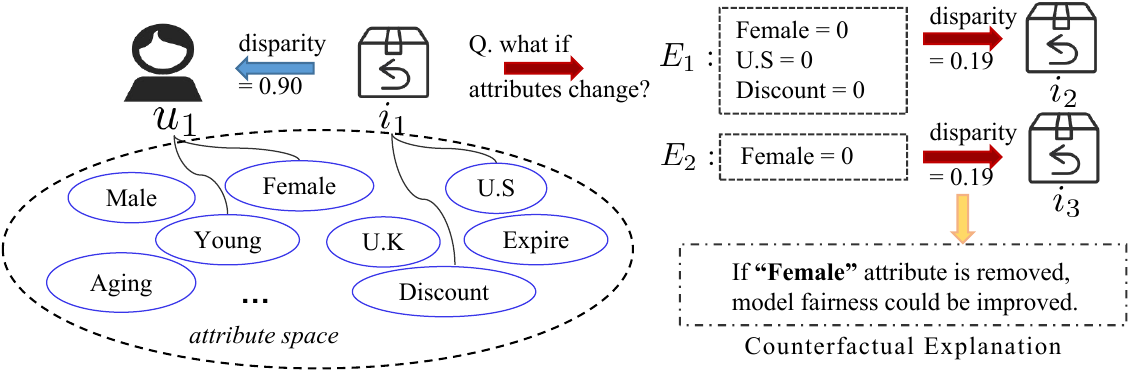}
    \caption{Toy example of inferring the attribute-level counterfactual explanation for fairness.}
    \label{fig:toy}
\end{figure}

Following the above intuition, we propose to generate attribute-level counterfactual explanations for fairness from a given HIN. 
We posit a novel definition of counterfactual explanation for fairness - \emph{a minimal set of attributes from the HIN that changes model fairness disparity}.
We use a toy example in Figure~\ref{fig:toy} to illustrate our idea.
Given a recommendation $i_1$ for the user $u_1$ and an external HIN carrying their attributes, we want to know why $i_1$ causes discrimination in recommendation results.
The counterfactual explanation performs ``what-if'' reasoning by altering the attributes of $u_1$ and $i_1$ and checking the fairness of the recommendation results.
Both $E_1$ and $E_2$ are valid candidate explanations since they alter fairness disparities of recommendations (i.e., $i_2$, $i_3$) from 0.90 to 0.19.  
To determine which attributes are the primary reason for unfairness, the counterfactual explanation will uncover the minimal attribute changes, i.e., $E_2$, instead of utilizing attribute combinations in $E_1$. 
Thus, we could infer $E_2$ is the most vital reason for model unfairness. 
Besides, since a counterfactual explanation $E_2$ is minimal, it only reveals the essential attributes (i.e., ``Female'') that effectively explain unfairness, while discarding the irrelevant (i.e., pseudo) explanations, i.e., ``U.S'' and ``Discount'' in $E_1$.

We therefore propose a novel \emph{Counterfactual Explanation for Fairness (CFairER)} within an off-policy reinforcement learning environment to find optimal attribute-level counterfactual explanations.
Particularly, we focus on generating attribute-level counterfactual explanations for item exposure unfairness to promote the fair allocation of user-preferred but less exposed items.
Note that the proposed approach is general and can be utilized in different recommendation scenarios that involve different fairness definitions.
Specifically, we use a reinforcement learning agent in \emph{CFairER} to optimize a fairness explanation policy by uniformly exploring candidate counterfactuals from a given HIN.
We also devise attentive action pruning over the HIN to reduce the search space of reinforcement learning.
Finally, our \emph{CFairER} optimizes the explanation policy using an unbiased counterfactual risk minimization objective, resulting in accurate attribute-level counterfactual explanations for fairness.
The contributions of this work are:
\begin{itemize}
    \item We make the first attempt to leverage rich attributes in a Heterogeneous Information Network to offer attribute-level counterfactual explanations for recommendation fairness.
    \item We propose an off-policy learning framework to identify optimal counterfactual explanations, 
    which is guided by an attentive action pruning to reduce the search space. 
    \item We devise a counterfactual risk minimization for off-policy correction, so as to achieve unbiased policy optimization.
    \item Comprehensive experiments show the superiority of our method in generating trustworthy explanations for fairness while preserving satisfactory recommendation performance. 
     
\end{itemize}

\section{Related Work\label{related work}}

\subsection{Fairness Explanation for Recommendation}
Recommender systems have long dealt with major concerns of recommendation unfairness, which profoundly harm user satisfaction~\cite{li2021user,fu2020fairness} and stakeholder benefits~\cite{beutel2019fairness,ge2022toward,li2022fairness}.
Recent works on fairness-aware recommendation mainly discuss two primary topics, i.e., user-side fairness~\cite{yao2017beyond,chen2018investigating,li2021towards,li2021user,fu2020fairness} and item-side fairness~\cite{abdollahpouri2017controlling,diaz2020evaluating,ge2021towards,liu2020balancing}.
User-side fairness concerns whether the recommendation is fair to different users/user groups, e.g., retaining equivalent accuracy or recommendation explainability.
Relevant approaches attribute the causes of user-side unfairness to discrimination factors, such as sensitive features (e.g., gender~\cite{yao2017beyond,chen2018investigating}, race~\cite{li2021towards}) and user inactiveness~\cite{li2021user,fu2020fairness}, etc. 
They mainly propose fairness metrics to constraint recommendation models (e.g., collaborative filtering~\cite{yao2017beyond}) to produce fair recommendations.
For example, Yao et al.~\cite{yao2017beyond} study the unfairness of collaborative filtering (CF)-based recommenders on gender-imbalanced data.
They propose four metrics to assess different types of fairness, then add these metrics as constraints to the CF model learning objective to produce fair recommendations.
Li et al.~\cite{li2021user} investigate the unfair recommendation between active and inactive user groups, and provide a re-ranking approach to mitigate the activity unfairness by adding constraints over evaluation metrics of ranking.
As modern content providers are more concerned about user privacy, it is generally not easy to access sensitive user features for the recommendation~\cite{resheff2018privacy}.
Meanwhile, users often prefer not to disclose personal information that raises discrimination~\cite{beigi2020privacy}.
Thus, another topic of item-side fairness-aware recommendation~\cite{abdollahpouri2017controlling,diaz2020evaluating,ge2021towards,liu2020balancing} is interested in examining whether the recommendation treats items fairly, e.g., similar ranking prediction errors for different items, fair allocations of exposure to each item.
For instance, 
Abdollahpouri et al. ~\cite{abdollahpouri2017controlling} address item exposure unfairness in learning-to-rank (LTR) recommenders.
They include a fairness regularization term in the LTR objective function, which controls the recommendations favored toward popular items.
Ge et al.~\cite{ge2021towards} consider the dynamic fairness of item exposure due to changing group labels of items.
They calculate the item exposure unfairness with a fairness-related cost function.
The cost function is merged into a Markov Decision Process to capture the dynamic item exposure for recommendations.
Liu et al.~\cite{liu2020balancing} focus on item exposure unfairness in interactive recommender systems (IRS).
They propose a reinforcement learning method to maintain a long-term balance between accuracy and exposure fairness in IRS.

Despite the great efforts, fairness-aware recommendations mitigate user and item unfairness in a black-box manner but do not explain why the unfairness appears.
Understanding the ``why'' is desirable for both model transparency~\cite{li2022fairness} and facilitates data curation to remove unfair factors~\cite{wang2022survey2}. 
Limited pioneering studies are conducted to explain fairness. 
Begley et al.~\cite{begley2020explainability} estimate Shapley values of input features to search which features contribute more to the model unfairness.
Ge et al.~\cite{DBLP:conf/sigir/GeTZXL0FGLZ22} develop an explainable fairness model for recommendation to explain which item aspects influence item exposure fairness.
They perform perturbations on item aspect scores, then apply perturbed aspect scores on two pre-defined matrices to observe fairness changes.
These prior efforts suffer from major limitations:
1) The high computational burden caused by the large-scale search space and the greedy nature of the explanation search process. 
2) They generate explanations by feature~\cite{begley2020explainability} or aspect~\cite{DBLP:conf/sigir/GeTZXL0FGLZ22} scores, which do not apply to discrete attributes such as gender and race.
Our work conducts counterfactual reasoning to seek minimal sets of attributes as explanations.
We also reduce the large search space by attentive action pruning in the off-policy learning environment. 
Meanwhile, we consider explaining recommendation unfairness based on attributes from a Heterogeneous Information Network, which is expected to be wildly applicable.

\subsection{Heterogeneous Information Network in Recommendation}
Heterogeneous Information Network (HIN) is a powerful structure that allows for the heterogeneity of its recorded data, i.e., various types of attributes, thus providing rich information to empower recommendations~\cite{wang2020joint,wang2022causal}.
HINs have been wildly adopted in recommendation models to boost performance; 
representative works cover context-based filtering (e.g., SemRec~\cite{shi2019semrec}, HERec~\cite{shi2018heterogeneous}) and knowledge-based systems (e.g., MCrec~\cite{hu2018leveraging}, HAN~\cite{wang2019heterogeneous}). 
For instance, HERec~\cite{shi2018heterogeneous} embeds meta-paths within a HIN as dense vectors, then fuses these HIN embeddings with user and item embeddings to augment the semantic information for recommendations. 
MCrec~\cite{hu2018leveraging} leverages a deep neural network to model meta-path-based contextual embeddings and propagates the context to user and item representations with a co-attention mechanism. 
Those recommendation models observe promising improvements by augmenting contextual and semantic information given by HINs. 
Despite the great efforts, prior works do not consider using the HIN to explain unfair factors in recommendations. 
Novel to this work, we first attempt to leverage rich attributes in a HIN to provide counterfactual explanations for item exposure fairness.

\subsection{Counterfactual Explanation} 

Counterfactual explanations have been considered as satisfactory explanations~\cite{woodwardMakingThingsHappen2004,li2022deep} and elicit causal reasoning in humans~\cite{byrne2007rational,yu2022semantics}.
Works on counterfactual explanations have been proposed very recently to improve the explainability of recommendations.
Xiong et al.~\cite{xiong2021counterfactual} propose a constrained feature perturbation on item features and consider the perturbed item features as explanations for ranking results.
Ghazimatin et al.~\cite{ghazimatin2020prince} perform random walks over a Heterogeneous Information Network to look for minimal sets of user action edges (e.g., click) that change the PageRank scores.
Tran et al.~\cite{tran2021counterfactual} identify minimal sets of user actions that update the parameters of neural models.
Our work differs from prior works on counterfactual explanations by two key points: 
1) In terms of problem definition, they generate counterfactual explanations to explain user behaviors (e.g., click~\cite{ghazimatin2020prince,tran2021counterfactual} ) or recommendation (e.g., ranking~\cite{xiong2021counterfactual}) results.
Our method generates counterfactual explanations to explain which attributes affect recommendation fairness.
2) In terms of technique, our method formulates counterfactual reasoning as reinforcement learning, which can deal with ever-changing item exposure unfairness.

\section{Preliminary}\label{pre} 
We first introduce the Heterogeneous Information Network that offers real-world attributes for fairness explanation learning.
We then give the key terminologies, including fairness disparity evaluation and counterfactual explanation for fairness.

\subsection{Heterogeneous Information Network}\label{sec:notion}
Creating fairness explanations requires auxiliary attributes containing possible factors (e.g., user gender) that affect recommendation fairness (cf. Figure~\ref{fig:toy}). 
Heterogeneous Information Network (HIN) has shown its power in modeling various types of attributes, e.g., user social relations, item brand.
In particular, suppose we have the logged data that records users’ historical behaviors (e.g., clicks) in the recommendation scenario.
Let $\mathcal{U} \in \mathbb{R}^{M}$, $\mathcal{I} \in \mathbb{R}^{N}$ denote the sets of users and items, respectively.
We can define a user-item interaction matrix $\boldsymbol{Y}=\left\{y_{uv} \mid u \in \mathcal{U}, v \in \mathcal{I}\right\}$ according to the logged data.
We also have additional attributes from external resources that profile users and items, e.g., users' genders, items' genres.
The connections between all attributes and users/items are absorbed in the relation set $\mathcal{E}$.
Those attributes, with their connections with user-item interactions, are uniformly formulated as a HIN.
Formally, a HIN is defined as $\mathcal{G}=(\mathcal{V}^\prime,\mathcal{E}^\prime)$, where $\mathcal{V}^\prime=\mathcal{U} \cup \mathcal{I} \cup \mathcal{V}_U \cup \mathcal{V}_I$, and $\mathcal{E}^\prime= \{\mathbb{I}({y_{uv})}\} \cup \mathcal{E}$.
$\mathbb{I}(\cdot)$ is an edge indicator that denotes the observed edge between user $u$ and item $v$ when $y_{uv} \in \boldsymbol{Y}=1$.
$\mathcal{V}_U$ and $\mathcal{V}_I$ are attribute sets for users and items, respectively.
Each node $n \in \mathcal{V}^\prime$ and each edge $e \in \mathcal{E}^\prime$ are mapped into specific types through node type mapping function: $\phi: \mathcal{V}^\prime \to \mathcal{K}$ and edge type mapping function: $\psi: \mathcal{E}^\prime \to \mathcal{J}$.
$\mathcal{G}$ maintain heterogeneity, i.e., $|\mathcal{K}|+|\mathcal{J}| > 2 $.

\subsection{Fairness Disparity}
We consider explaining the item exposure (un)fairness in recommendations.
We first split items in historical user-item interactions into head-tailed (i.e., popular) group $G_{0}$ the long-tailed group $G_{1}$~\footnote{Following~\cite{DBLP:conf/sigir/GeTZXL0FGLZ22}, we consider the top 20\% items with the most frequent interactions with users as $G_{0}$, while the remaining 80\% belongs to $G_{1}$.}.
Following previous works~\cite{DBLP:conf/sigir/GeTZXL0FGLZ22,ge2021towards}, we use demographic parity (DP) and exact-K (EK) defined on item subgroups to measure whether a recommendation result is fair.
In particular, DP requires that each item has the same likelihood of being classified into $G_{0}$ and $G_{1}$.
EK regulates the item exposure across each subgroup to remain statistically indistinguishable from a given maximum $\alpha$.
By evaluating the deviation of recommendation results from the two fairness criteria, we can calculate the fairness disparity, i.e., to what extent the recommendation model is unfair. 
Formally, giving a recommendation result $H_{u, K}$, the fairness disparity $\Delta(H_{u, K})$ of $H_{u, K}$ is: 
\begin{equation}\label{eq:disparity}
\begin{array}{ll}
     \Delta(H_{u, K})=\left|\Psi_{D P}\right|+\lambda \left|\Psi_{E K}\right| \text{, } \\
     \Psi_{D P}=\left|G_{1}\right| \cdot \text { Exposure }\left(G_{0} \mid H_{u, K}\right)-\left|G_{0}\right| \cdot \text { Exposure }\left(G_{1} \mid H_{u, K}\right), \\
     \Psi_{E K}= \alpha \cdot \text { Exposure }\left(G_{0} \mid H_{u, K}\right)-\text { Exposure }\left(G_{1} \mid H_{u, K}\right)
\end{array}
\end{equation}
where $\Delta(\cdot)$ is the fairness disparity metric that quantifies model fairness status.
$\lambda$ is the trade-off parameter between DP and EK.
$\text{Exposure}\left(G_{j} \mid H_{u, K}\right)$ is the item exposure number of $H_{u, K}$ within $G_j$ w.r.t. $j \in\{0,1\}$.

\subsection{Counterfactual Explanation for Fairness}\label{sec:counterfactual fairness} 

This work aims to generate attribute-level counterfactual explanations for item exposure fairness.
In particular, we aim to find the ``minimal'' changes in attributes that reduce the fairness disparity (cf. Eq.~\eqref{eq:disparity}) of item exposure.
Formally, given historical user-item interaction $\boldsymbol{Y}=\left\{y_{uv} \mid u \in \mathcal{U}, v \in \mathcal{I}\right\}$, and user attribute set $\mathcal{V}_U$ and item attribute set $\mathcal{V}_I$ extracted from an external Heterogeneous Information Network (HIN) $\mathcal{G}=(\mathcal{V}^\prime,\mathcal{E}^\prime)$.
Suppose there exists a recommendation model that produces the recommendation result $H_{u, K}$ for user $u$. 
Given all user-item pairs $(u,v)$ in $H_{u, K}$,
our goal is to find a minimal attributes set $\mathcal{V}^{*} \subseteq \{ \{e_u, e_v\} \mid (u, e_u), (v, e_v) \in \mathcal{E}^\prime, e_u \in \mathcal{V}_U, e_v \in \mathcal{V}_I\}$.
Each attribute in $\mathcal{V}^{*}$ is an attribute entity from HIN $\mathcal{G}$, e.g., user's gender, item's genre. 
With a minimal set of $\mathcal{V}^{*}$, the counterfactual reasoning pursues to answer: what the fairness disparity would be, if $\mathcal{V}^{*}$ is applied to the recommendation model.
$\mathcal{V}^{*}$ is recognized as a valid \emph{counterfactual explanation for fairness}, if after applied $\mathcal{V}^{*}$, the fairness disparity of the intervened recommendation result $\Delta(H_{u, K}^{cf})$ reduced compared with original $\Delta(H_{u, K})$.
In addition, $\mathcal{V}^{*}$ is \emph{minimal} such that there is no smaller set $\mathcal{V}^{*^{\prime}}\in \mathcal{G}$ satisfying $|\mathcal{V}^{*^{\prime}}| < |\mathcal{V}^{*}|$ when $\mathcal{V}^{*^{\prime}}$ is also valid.

\section{The \emph{CFairER} Framework}\label{sec:framework}

\begin{figure}[htbp]
\centering
\includegraphics[width=0.8\textwidth]{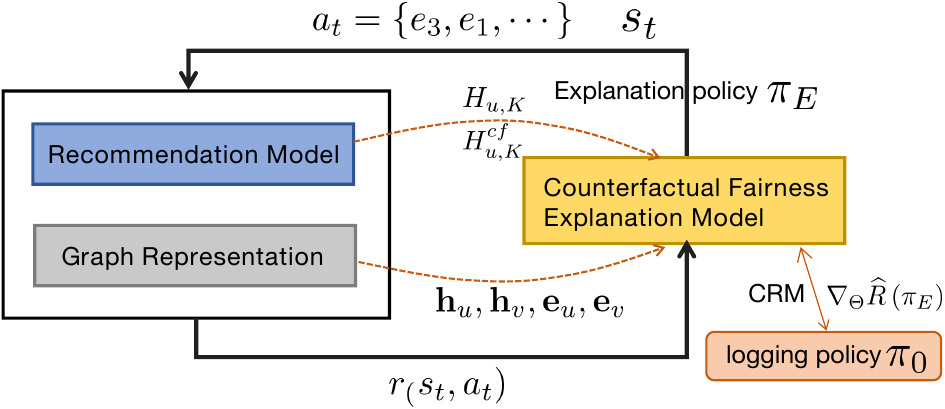}
\caption{The proposed \emph{CFairER} framework.
}
\label{fig:frame}
\end{figure}

We now introduce the framework of our \emph{Counterfactual Explanation for Fairness (CFairER)}.
As shown in Figure~\ref{fig:frame}, \emph{CFairER} devises three major components: 
1) graph representation module embeds users, items, and attributes among HIN as embedding vectors; 
2) recommendation model learns user and item latent factors to produce recommendation results and 
3) our proposed counterfactual fairness explanation (CFE) model assisted by the graph representation module and the recommendation model to conduct counterfactual reasoning.
This section discusses how the CFE model collaborates with the other two components, then introduces the graph representation module and the recommendation model.
We will elaborate on our proposed CFE model in the next section.

\subsection{Counterfactual Fairness Explanation Model}
As shown in Figure~\ref{fig:frame}, our CFE model is crafted within an off-policy learning environment, in which an explanation policy $\pi_E$ is optimized to produce attribute-level counterfactual explanations for fairness.
At each state $s_t$, $\pi_E$ produces actions $a_t$ absorbing user and item attributes as potential counterfactual explanations.
These actions are committed to the recommendation model and graph representation module to produce the reward $r(s_t, a_t)$ for optimizing $\pi_E$. 
Specifically, the graph representation module provides dense vectors $\mathbf{h}_{u}$, $\mathbf{h}_v$, $\mathbf{e}_{u}$ and $\mathbf{e}_{v}$ as user, item, user attribute and item attribute embeddings, respectively.
Those embeddings are used in the state representation learning (i.e., learn $s_t$) and attentive action pruning (i.e., select $a_t$) in our CFE model.
Moreover, the attribute embeddings are fused with user or item latent factors learned by the recommendation model to explore the model fairness change.
In particular, the fused embeddings of users and items are used to predict the intervened recommendation result $H_{u, K}^{cf}$.
By comparing the fairness disparity (cf. Eq.~\eqref{eq:disparity}) difference between $H_{u, K}^{cf}$ and the original recommendation $H_{u, K}$, we determine the reward $r(s_t, a_t)$ to optimize $\pi_E$, accordingly. 
The reward $r(s_t, a_t)$ measures whether the current attribute (i.e., action) is a feasible fairness explanation responsible for the fairness change. 
Finally, $\pi_{E}$ is optimized with a counterfactual risk minimization (CRM) objective $\nabla_{\Theta} \widehat{R}\left(\pi_E\right)$ to balance the distribution discrepancy from the logging policy $\pi_0$.

\subsection{Graph Representation Module}\label{sec:graph representation}
Our graph representation module conducts heterogeneous graph representation learning to produce dense vectors of users, items, and attributes among the HIN.
Compared with homogeneous graph learning such as GraphSage~\cite{hamilton2017inductive}, our graph representation injects both node and edge heterogeneity to preserve the complex structure of the HIN.
In particular, we include two weight matrices to specify varying weights of different node and edge types.

In the following, we present the graph learning for user embedding $\mathbf{h}_{u}$.
The embeddings of $\mathbf{h}_{v}$, $\mathbf{e}_u$ and $\mathbf{e}_v$ can be obtained analogously by replacing nodes and node types while computations. 
Specifically, we first use Multi-OneHot~\cite{zhang2019stylistic} to initialize node embeddings at the $0$-th layer, in which $u$'s embedding is denoted by $\mathbf{h}_u^{0}$.
Then, at each layer $l$, user embedding $\mathbf{h}_{u}^{l}$ is given by aggregating node $u$'s neighbor information w.r.t. different node and edge types:
\begin{equation}\label{eq:each layer}
\small
   \mathbf{h}_{u}^{l}=\sigma\left(\operatorname{concat }\left[\mathbf{W}_{\phi(u)}^{l} \operatorname{D}_{p}\left[\mathbf{h}_{u}^{l-1}\right],  \frac{\mathbf{W}_{\psi(e)}^{l}}{\left|\mathcal{N}_{\psi(e)}(u)\right|} \sum_{u^\prime \in \mathcal{N}_{\psi(e)}(u)} \operatorname{D}_{p}\left[\mathbf{h}_{u^\prime}^{l-1}\right] \right]+b^{l}\right)
\end{equation}
where $\sigma(\cdot)$ is LeakyReLU~\cite{xu2020reluplex} activation function and $\operatorname{concat}(\cdot)$ is the concatenation operator. 
$\operatorname{D}_{p}[\cdot]$ is a random dropout with probability $p$ applied to its argument vector.
$\mathbf{h}_{u}^{l-1}$ is $u$'s embedding at layer $l-1$.
$\mathcal{N}_{\psi(e)}(u)=\left\{u^{\prime} \mid \left(u, e, u^{\prime}\right) \in \mathcal{G} \right\}$ is a set of nodes connected with user node $u$ through edge type $\psi(e)$.
The additionally dotted two weight matrices, i.e., node-type matrix $\mathbf{W}_{\phi(u)}^{l}$ and edge-type matrix $\mathbf{W}_{\psi(e)}^{l}$, are defined based on the importance of each type $\phi(u)$ and $\psi(e)$.
$b^{l}$ is an optional bias.

With Eq~\eqref{eq:each layer}, we obtain $u$'s embedding $\mathbf{h}_{u}^{l}$ at each layer $l \in \{1,\cdots, L\}$.
We then adopt layer-aggregation~\cite{xu2018representation} to concatenate $u$'s embeddings at all layers into a single vector, i.e., $\mathbf{h}_u=\mathbf{h}_u^{(1)} + \cdots + \mathbf{h}_u^{(L)}$.
Finally, we have user node $u$'s embedding $\mathbf{h}_u$ through aggregation.
The item embedding $\mathbf{h}_{v}$, user attribute embedding $\mathbf{e}_u$ and item attribute embedding $\mathbf{e}_v$ can be calculated analogously.

\subsection{Recommendation Model}\label{sec:rec}
The recommendation model $f_R$ is initialized using user-item interaction matrix $\boldsymbol{Y}$ to produce the Top-$K$ recommendation result $H_{u, K}$ for all users.
Here, we employ a linear and simple matrix factorization (MF)~\cite{rendle2012bpr} as the recommendation model $f_R$. 
Particularly, MF initializes IDs of users and items as latent factors, and uses the inner product of user and item latent factors as the predictive function:
\begin{equation}\label{eq:uv}
    f_R(u,v)=\boldsymbol{U}_{u}^{\top} \boldsymbol{V}_{v}
\end{equation}
where $\boldsymbol{U}_{u}$ and $\boldsymbol{V}_{v}$ denote $d$-dimensional latent factors for user $u$ and item $v$, respectively. 
We use the cross-entropy~\cite{zhang2018generalized} loss to define the objective function of the recommendation model: 
\begin{equation}
\label{eq:recommendation model}
\mathcal{L}_R = -\sum_{u, v, y_{uv} \in \boldsymbol{Y}} y_{uv} \log f_R({u,v})+\left(1-y_{uv}\right) \log \left(1-f_R({u,v})\right) 
\end{equation}
After optimizing the loss function $\mathcal{L}_R$, we can use the trained user and item latent factors (i.e., $\boldsymbol{U}$, $\boldsymbol{V}$) to produce the original Top-$K$ recommendation lists $H_{u, K}$ for all users $u \in \mathcal{U}$.

\section{Reinforcement Learning for Counterfactual Fairness Explanation
}\label{sec:CFE}

\begin{figure}[htbp]
\centering
\includegraphics[width=0.8\textwidth]{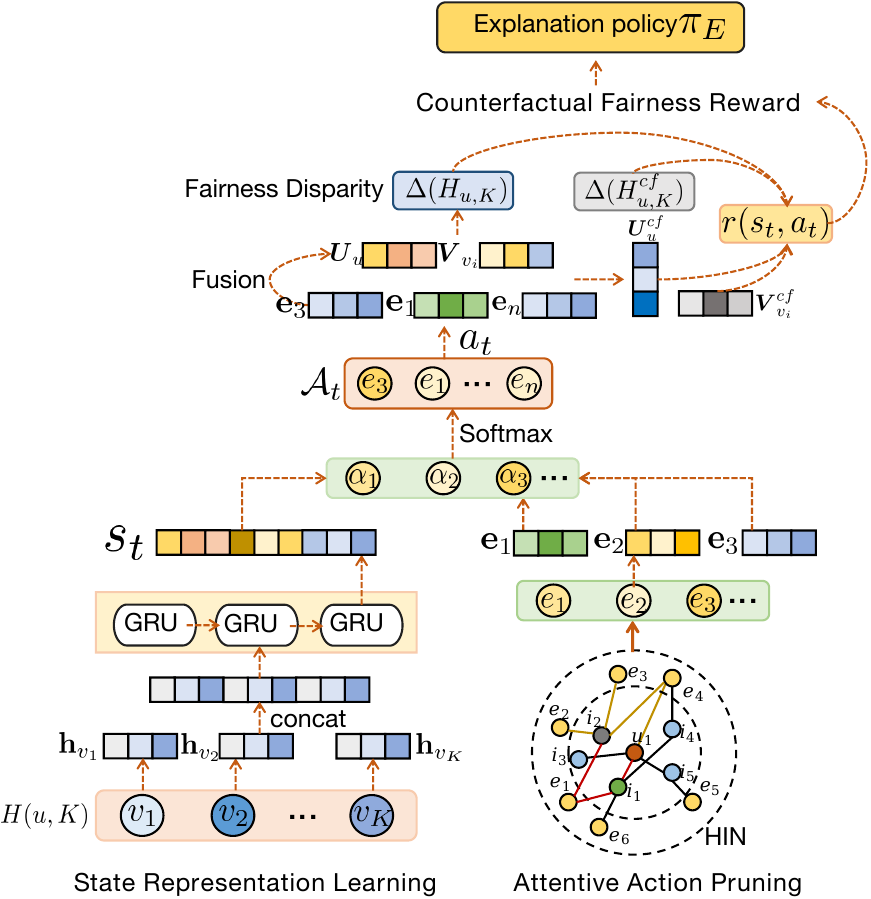}
\caption{Counterfactual Fairness Explanation (CFE) Model.
}
\label{fig:ct_model}
\end{figure}

We put forward our counterfactual fairness explanation (CFE) model (cf. Figure~\ref{fig:ct_model}), assisted by graph representation module and recommendation model, to generate explanation policy $\pi_E$ for item exposure fairness. 
The explanation policy $\pi_E$ is optimized within off-policy learning to adaptively learn attributes responsible for fairness changes.
In the following, we first introduce off-policy learning for our CFE model.
Then we detail each key element in the off-policy learning and give unbiased policy optimization.

\subsection{Explaining as Off-policy Learning }\label{sec:CEF}

We cast our CFE model in an off-policy learning environment, which is formulated as Markov Decision Process (MDP).
The MDP is provided with a static logged dataset generated by a logging policy $\pi_0$~\footnote{We adopt the uniform-based logging policy as $\pi_0$. It samples attributes as actions from the attribute space with the probability of $\pi_0(a_t \mid s_t)=\frac{1}{|\mathcal{V}_U+\mathcal{V}_I|}$.}.
The logging policy $\pi_0$ collects trajectories by uniformly sampling actions from the user and item attribute space. 
We use the off-policy learning to optimize an explanation (i.e., target) policy $\pi_E$ by approximating the counterfactual rewards of state-action pairs from all timestamps, wherein the logging policy $\pi_0$ is employed for exploration while the target policy $\pi_E$ is utilized for decision-making. 
In the off-policy setting, 
the explanation policy $\pi_E$ does not require following the original pace of the logging policy $\pi_0$.
As a result, $\pi_E$ is able to explore the counterfactual region, i.e., those actions that haven't been taken by the previous agent using $\pi_0$.
Formally, at each timestamp $t \in \{1,\cdots,T\}$ of MDP, the explanation policy $\pi_E(a_t|s_t)$ selects an action (i.e., a candidate attribute) $a_t \in \mathcal{A}_t$ conditioning on the user state $s_t \in \mathcal{S}$, and receives counterfactual reward $r(s_t, a_t) \in \mathcal{R}$ for this particular state-action pair.
Then the current state transits to the next state $s_{t+1}$ with transition probability of $\mathbb{P}\left(s_{t+1}\mid s_{t}, a_{t}\right)\in \mathcal{P}$.  
The whole MDP has the key elements:
\begin{itemize}
    \item $\mathcal{S}$ is a finite set of states $\{s_t \mid t\in [1,\cdots, T]\}$. Each state $s_t$ is transformed into dense vectors (i.e., embeddings) by our \emph{state representation learning} (cf. Section~\ref{sec:state representation}).
    
    \item $\mathcal{A}_t$ is a finite set of actions (i.e., attributes) available at $s_t$. $\mathcal{A}_{t}$ is select from attributes $\mathcal{V}_t \in \mathcal{G}$ by our \emph{attentive action pruning} (cf. Section~\ref{sec:action pruning}) to reduce the search space. 

    \item $\mathcal{P}: \mathcal{S} \times \mathcal{A} \rightarrow \mathcal{S}$ is the state transition, which absorbs transition probabilities of the current states to the next states.  
    Given action $a_t$ at state $s_t$, the transition to the next state $s_{t+1}$ is determined as
    $\mathbb{P}\left(s_{t+1}\mid s_{t}, a_{t}\right)\in \mathcal{P} =1$.
    
    \item $\mathcal{R}: \mathcal{S} \rightarrow \mathcal{R}$ is the counterfactual reward measures whether a deployed action (i.e., an attribute) is a valid counterfactual explanation for fairness. $\mathcal{R}$ is used to guide the explanation policy learning and is defined in Section~\ref{sec:reward}.
\end{itemize}
We now introduce the implementation of each key component.

\subsubsection{State Representation Learning.}\label{sec:state representation}
The state $\mathcal{S}$ describes target users and their recommendation lists from the recommendation model.
Formally, at step $t$, the state $s_t$ for a user $u$ is defined as $s_t=(u, H(u,K))$, where $u \in \mathcal{U}$ is a target user and $H(u,K)$ is the recommendation produced by $f_R$.
The initial state $s_0$ is $(u, v)$ and $v$ is the interacted item of $u$, i.e., $y_{uv} \in \boldsymbol{Y}=1$.
Our state representation learning maps user state $s_t=(u, H(u,K))$ into dense vectors for latter explanation policy learning. 
Specifically, given $s_t$ that absorbs current user $u$ and its recommendation $H(u,K)=\{v_1,v_2,...,v_K\}$.
We first acquire the embedding $\mathbf{h}_{v_k}$ of each item $v_k \in H(u,K)$ from our graph representation module.
The state $s_t$ then receives the concatenated item embeddings (i.e., $\operatorname{concat}\left[\mathbf{h}_{v_k}|\forall v_k \in H(u,K)\right]$) to update its representation.  
Considering states within $\mathcal{S}$ have sequential patterns~\cite{10.1145/3485447.3512072}, 
we resort to Recurrent Neural Networks (RNN) with a gated recurrent unit (GRU)~\cite{dey2017gate} to capture the sequential state trajectory. 
We firstly initialize the state representation $s_0$ with an initial distribution $s_{0} \sim \rho_{0}$ 
\footnote{In our experiment, we used a fixed initial state distribution, where $s_{0} = 0 \in \mathbb{R}^{d}$}.
Then we learn state representation $s_t$ through the recurrent cell:
\begin{equation}\label{eq:gru}
    \begin{aligned}
\mathbf{u}_{t} &=\sigma_{g}\left(\mathbf{W}_{1} \operatorname{concat}\left[\mathbf{h}_{v_k}|\forall v_k \in H(u,K)\right]+\mathbf{U}_{1} s_{t-1}+b_1\right) \\
\mathbf{r}_{t} &=\sigma_{g}\left(\mathbf{W}_2 \operatorname{concat}\left[\mathbf{h}_{v_k}|\forall v_k \in H(u,K)\right]+\mathbf{U}_{2} s_{t-1}+{b}_{2}\right) \\
\hat{s}_{t} &=\sigma_{h}\left(\mathbf{W}_{3} \operatorname{concat}\left[\mathbf{h}_{v_k}|\forall v_k \in H(u,K)\right]+\mathbf{U}_{3}\left(\mathbf{r}_{t} \cdot s_{t-1}\right)+{b}_3\right) \\
s_{t} &=\left(1-\mathbf{u}_{t}\right) \cdot s_{t-1}+\mathbf{u}_{t} \odot \hat{s}_{t}
    \end{aligned}
\end{equation}
where $\mathbf{u}_{t}$ and $\mathbf{r}_{t}$ denote the update gate and reset gate vector generated by GRU and $\odot$ is the element-wise product operator.
$\mathbf{W}_{i}$, $\mathbf{U}_{i}$ are weight matrices and ${b}_{i}$ is the bias vector.  
Finally, $s_{t}$ serves as the state representation at time step $t$.

\subsubsection{Attentive Action Pruning.}\label{sec:action pruning}

Our attentive action pruning is designed to reduce the action search space by specifying the varying importance of actions for each state.
As a result, the sample efficiency can be largely increased by filtering out irrelevant actions to promote an efficient action search. 
In our method, actions are defined as candidate attributes selected from a given HIN that potentially impact the model fairness. 
In particular, given state $s_t=(u, H(u,K))$, we can distill a set of attributes $\mathcal{V}_t$ of the current user $u$ and items $v \in H(u,K)$ from the HIN.
Intuitively, we can directly use $\mathcal{V}_{t}$ as candidate actions for state $s_t$. 
However, the user and item attribute amount of the HIN would be huge, resulting in a large search space that terribly degrades the learning efficiency~\cite{10.1145/3477495.3532021}.
Thus, we propose an attentive action pruning based on attention mechanism~\cite{vaswani2017attention} to select important candidate actions for each state.
Formally, given the embedding $\mathbf{e}_{i}$ for an attribute $i \in \mathcal{V}_t$ from Eq.~\eqref{eq:each layer}, and the state representation $s_{t}$ from Eq.~\eqref{eq:gru}, the attention score $\alpha_{i}$ of attribute $i$ is:
\begin{equation}\label{eq:candidate_attention}
\alpha_{i}=\operatorname{ReLU}\left(\mathbf{W}_{s} s_{t}+\mathbf{W}_{h} \mathbf{e}_{i}+{b}\right) 
\end{equation}
where $\mathbf{W}_{s}$ and $\mathbf{W}_{h}$ are two weight matrices and ${b}$ is the bias vector.

We then normalize attentive scores of all attributes in $\mathcal{V}_{t}$ and select attributes with $n$-top attention scores into $\mathcal{A}_t$:
\begin{equation}\label{eq:candidate}
\mathcal{A}_t=\left\{i \mid i \in \text{Top-n} \left[\frac{\exp \left(\alpha_{i}\right)}{\sum_{i^\prime \in \mathcal{V}_{t}} \exp \left(\alpha_{i^{\prime}}\right)}\right]  \text { and } i \in \mathcal{V}_{t} \right\}
\end{equation}
where $n$ is the candidate size.
To the end, our candidate set $\mathcal{A}_t$ is of high sample efficiency since it filters out irrelevant attributes while dynamically adapting to the user state shift.

\subsubsection{Counterfactual Reward Definition}\label{sec:reward}
The counterfactual reward $r(s_t, a_t) \in \mathcal{R}$ measures whether a deployed action $a_t \in \mathcal{A}_t$ is a valid counterfactual explanation for fairness at the current state $s_t$.
In particular, the reward is defined based on two criteria:
1) \textit{Rationality}~\cite{verma2020counterfactual}: deploying action (i.e., attribute) $a_t$ should cause the reduction of fairness disparity regarding the item exposure fairness.
The fairness disparity change is measured by the fairness disparity difference between the recommendation result before (i.e., $\Delta(H_{u, K})$) and after (i.e., $\Delta(H_{u, K}^{cf})$) fusing the action $a_t$ to the recommendation model $f_R$, i.e., $\Delta(H_{u, K})- \Delta(H_{u, K}^{cf})$.
2) \textit{Proximity}~\cite{DBLP:journals/corr/abs-2001-07417}: a counterfactual explanation is a minimal set of attributes that changes the fairness disparity.

For the \emph{Rationality}, we fuse the embedding of $a_t$ with user or item latent factors from the recommendation model to learn updated user and item latent vectors, so as to get the $\Delta(H_{u, K}^{cf})$.
Specifically, for a state $s_t=(u, H(u,K))$, the embedding $\mathbf{e}_{t}$ of action $a_t$ is fused to user latent factor $\boldsymbol{U}_{u}$ for user $u$ and item latent factors $\boldsymbol{V}_{v_i}$ for all items $v_i \in H(u,K)$ by a element-wise product fusion.
As a result, we can get the updated latent factors $\boldsymbol{U}_{u}^{cf}$ and $\boldsymbol{V}_{v}^{cf}$: 
\begin{equation}\label{eq:fusion}
\begin{aligned}
\boldsymbol{U}_{u}^{cf} & \leftarrow  \boldsymbol{U}_{u} \odot  \left\{\mathbf{e}_{t}  \mid \forall t \in [1, \cdots, T]\right\}, \text{if } a_t \in \mathcal{V}_U \\
\boldsymbol{V}_{v_i}^{cf} & \leftarrow  \boldsymbol{V}_{v_i} \odot \left\{\mathbf{e}_{t}   \mid \forall t \in [1, \cdots, T]\right\}, \text{if } a_t \in \mathcal{V}_I 
\end{aligned}
\end{equation}
where $\odot$ represents the element-wise product (a.k.a. Hadamard product).
$T$ is the total training iteration.
At the initial state of $t=0$, user and item latent factors $\boldsymbol{U}_{u}$ and $\boldsymbol{V}_{v}$ are learned form Eq~\eqref{eq:uv}.
Through Eq.~\eqref{eq:fusion}, the updated user and item latent vectors are then used to generate the intervened recommendation result $H_{u, K}^{cf}$.

For the \emph{Proximity}, we compute whether $a_t$ returns a minimal set of attributes that changes the recommendation model fairness. 
This is equal to regulating user and item latent factors before (i.e., $\boldsymbol{U}_{u}$, $\boldsymbol{V}_{v}$) and after (i.e., $\boldsymbol{U}_{u}^{cf}$, $\boldsymbol{V}_{v}^{cf}$) fusing $a_t$ be as similar as possible.

Based on the two criteria, the counterfactual reward can be defined as the following form:
\begin{equation}\label{eq:reward}
\small
r(s_t, a_t)=\left\{\begin{array}{l}1+\operatorname{dist} (\boldsymbol{U}_{u}, \boldsymbol{U}_{u}^{cf})+ \operatorname{dist} (\boldsymbol{V}_{v}, \boldsymbol{V}_{v}^{cf}), \text { if } \Delta(H_{u, K})- \Delta(H_{u, K}^{cf})  \geq \epsilon  \\ 
\operatorname{dist} (\boldsymbol{U}_{u}, \boldsymbol{U}_{u}^{cf})+ \operatorname{dist} (\boldsymbol{V}_{v}, \boldsymbol{V}_{v}^{cf}), \text { otherwise }\end{array}\right.
\end{equation}
where $\operatorname{dist}(\cdot)$ is the distance metric defined as cosine similarity~\cite{nguyen2010cosine}, i.e., $\operatorname{dist} (a,b)=\frac{\langle a, b\rangle}{\|a\| \| b\|}$. 
$\Delta(\cdot)$ is the fairness disparity evaluation metric defined in Eq.\eqref{eq:disparity}.
$\epsilon$ is the disparity change threshold that controls the model flexibility.

\subsection{Unbiased Policy Optimization}\label{sec:optimization}

Using state $s_t \in \mathcal{S}$ from Eq.~\eqref{eq:gru}, candidate action $a_t \in \mathcal{A}_t$ from Eq.~\eqref{eq:candidate}, and counterfactual reward $r(s_t, a_t)$ in Eq.~\eqref{eq:reward} for each timestamp $t$, 
the policy optimization seeks the explanation policy $\pi_E$ that maximizes the expected cumulative reward $R(\pi_E)$ over total iteration $T$.
Intuitively, we can directly use the policy gradient calculated on $R(\pi_E)$ to guide the optimization of $\pi_E$.
However, our policy optimization is conducted in the off-policy learning setting, in which $\pi_E$ holds different distribution from the logging policy $\pi_0$.
Directly optimizing $R(\pi_E)$ would result in a biased policy optimization~\cite{10.1145/3477495.3532021} due to the policy distribution discrepancy.
To this end, we additionally apply \textit{Counterfactual Risk Minimization} (CRM)~\cite{swaminathan2015counterfactual} to correct the discrepancy between $\pi_E$ and $\pi_0$.
In particular, CRM employs an Inverse Propensity Scoring (IPS)~\cite{williamson2014introduction} to explicitly estimate the distribution shift between $\pi_E$ and $\pi_{0}$.
After applying the CRM, we can alleviate the policy distribution bias by calculating the CRM-based expected cumulative reward $\widehat{R}(\pi_E)$:
\begin{equation}\label{eq:crm}
\small
\widehat{R}(\pi_E)
= \mathbb{E}_{\pi_E}\left[\sum_{t=0}^{T} \gamma^{t}  \frac{\pi_E\left(a_{t} \mid s_t\right)}{\pi_{0}\left(a_{t} \mid s_t\right)} r\left(s_{t}, a_{t}\right)\right]
\end{equation}
where $\frac{\pi_E(a_t \mid {s}_{t})}{\pi_0(a_t \mid {s}_{t})}$ is called the \textit{propensity score} for balancing the empirical risk estimated from the $\pi_0$.

Finally, the policy gradient of the explanation policy learning w.r.t. model parameter $\Theta$ is achieved by the REINFORCE~\cite{williams1992simple}:
\begin{equation}\label{eq:objective}
\nabla_{\Theta} 
\widehat{R}\left(\pi_E\right)=\frac{1}{T} \sum_{t=0}^{T}\gamma^{t}  \frac{\pi_E\left(a_{t} \mid s_t\right)}{\pi_{0}\left(a_{t} \mid s_t\right)} r\left(s_{t}, a_{t}\right) \nabla_{\Theta} \log \pi_E(a_t \mid s_t) 
\end{equation}
where $T$ is the total training iteration.
By optimizing the Eq.~\eqref{eq:objective}, the learned explanation policy $\pi_E$ generates minimal sets of attributes responsible for item exposure fairness changes, so as to find the true reasons leading to unfair recommendations.

\section{Experiments}
We conduct extensive experiments to evaluate the proposed \emph{CFairER} for explaining item exposure fairness in recommendations.
We aim to particularly answer the following research questions:
\begin{itemize}
    \item \textbf{RQ1.} Whether \emph{CFairER} produces attribute-level explanations that are faithful to explaining recommendation model fairness compared with existing approaches? 
    
    \item \textbf{RQ2.} Whether explanations provided by \emph{CFairER} 
    achieve better fairness-accuracy trade-off than other methods? 
    
    \item \textbf{RQ3.} Do different components (i.e., attentive action pruning, counterfactual risk minimization-based optimization) help \emph{CFairER} to achieve better sample efficiency and bias alleviation? How do hyper-parameters impact \emph{CFairER}?

\end{itemize}

\subsection{Experimental Setup}

\subsubsection{Datasets}
We use logged user behavior data from three datasets \texttt{Yelp}~\footnote{https://www.yelp.com/dataset/}, \texttt{Douban Movie}~\footnote{https://movie.douban.com/} and \texttt{Last-FM}~\footnote{https://github.com/librahu/HIN-Datasets-for-Recommendation-and-Network-Embedding} for evaluations.
Each dataset is considered as an independent benchmark for different tasks, i.e., business, movie and music recommendation tasks. 
The \texttt{Yelp} dataset records user ratings on local businesses and business compliment, category and city profiles.
The \texttt{Douban Movie} is a movie recommendation dataset that contains user group information and movie actor, director and type details.
The \texttt{Last-FM} contains music listening records of users and artist tags.
The details of both datasets are given in Table~\ref{tb:dataset}, which depicts statistics of user-item interactions, user-attribute and item-attribute relations.
All datasets constitute complex user-item interactions and diverse attributes, thus providing rich contextual information for fairness explanation learning.
Following previous works~\cite{10.1145/3477495.3532021,10.1145/3533725,wang2020joint}, we adopt a 10-core setting, i.e., retaining users and items with at least ten interactions for both datasets to ensure the dataset quality.
Meanwhile, we binarize the explicit rating data by interpreting ratings of 4 or higher as positive feedback, otherwise negative. 
Then, we sort the interacted items for each user based on the timestamp and split the chronological interaction list into train/test/valid sets with a proportion of 60\%/20\%/20\%.

\begin{table}[h]
\centering
\caption{Statistics of the datasets.}\label{tb:dataset}
\resizebox{0.7\textwidth}{!}{
\begin{tabular}{l|l|c|c|c}
\hline
\multicolumn{2}{c|}{Dataset}& \texttt{Yelp} & \texttt{Douban Movie}  & \texttt{Last-FM}  \\
\midrule
\multirow{3}{*}{\shortstack{User-Item\\ Interaction}}
& \#Users & 16,239 & 13,367 & 1,892
\\ & \#Items & 14,284 & 12,677  & 17,632
\\& \#Interactions & 198,397 & 1,068,278  & 92,834
\\ & \#Density & 0.086\% & 0.63\%   & 0.28\%
\\
\midrule
\multirow{3}{*}{\shortstack{User Item \\ Attributes}} & \#User Attributes & 16,250 & 16,120  &1,892 \\ 
&\#User-side Relations & 235,465  & 574,132  & 25,434  \\
&\#Item Attributes & 558 & 8,798  & 29,577  \\
&\#Item-side Relations &54,276 & 72,531  & 338,340  \\
\hline
\end{tabular}
}
\end{table}

\begin{figure}[htbp]
\centering
\begin{minipage}[t]{0.7\textwidth}
\centering
\includegraphics[width=\textwidth]{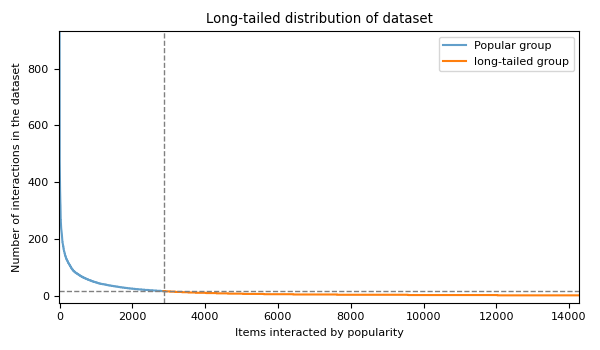}
\subcaption{\centering Interaction distribution in \texttt{Yelp}.}
\end{minipage}
\begin{minipage}[t]{0.7\textwidth}
\centering
\includegraphics[width=\textwidth]{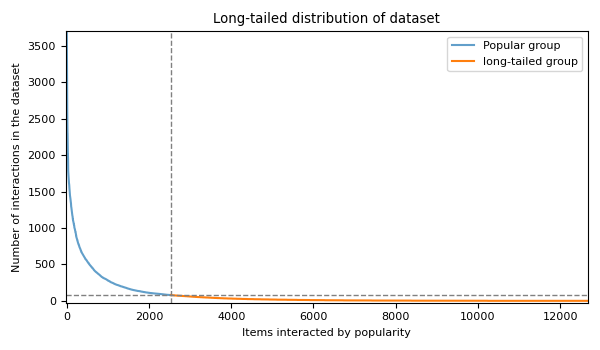}
\subcaption{\centering Interaction distribution in \texttt{Douban Movie}.}
\end{minipage}
\begin{minipage}[t]{0.7\textwidth}
\centering
\includegraphics[width=\textwidth]{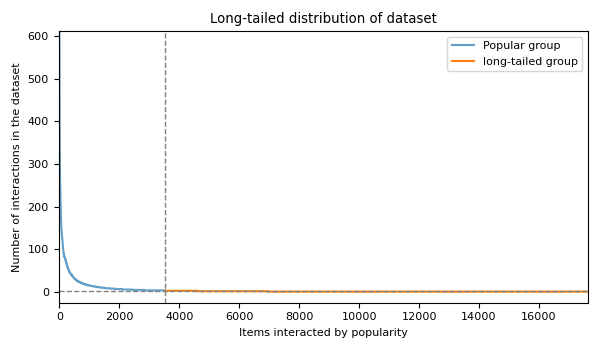}
\subcaption{\centering Interaction distribution in \texttt{Last-FM}.}
\end{minipage}
\caption{Long-tailed distribution of item exposure in \texttt{Yelp}, \texttt{Douban Movie}, and \texttt{Last-FM}.}
\label{fig:long-tail}
\end{figure}

We also study the long-tail distribution of user-item interactions in the three datasets.
We present the visualization results of the distribution of historical user-item interactions in the three datasets in Figure~\ref{fig:long-tail}.
Analyzing Figure~\ref{fig:long-tail}, we find that user-item interactions of both datasets are presented with a skewed distribution: the head-tailed distribution in the blue plot area and the long-tailed distribution in the yellow plot area. 
Besides, a small fraction of popular items account for most of the user interactions in both datasets, 
The skewed distribution would result in serious item exposure unfairness issues in recommendations, such as the well-known filter-bubble problem~\cite{nguyen2014exploring} and Matthew effect~\cite{perc2014matthew}.

\subsubsection{Baselines}
We adopt three heuristic approaches and two existing fairness-aware explainable recommendation methods as baselines.
In particular,

\begin{itemize}
    \item \textbf{RDExp}: We randomly select attributes from the attribute space for each user-item interaction and generate explanations based on the selected attributes. Note that the selected attributes can be both user and item attributes.
    \item \textbf{PopUser} and \textbf{PopItem}: We separately calculate the exposure number of attributes for each user-item interaction, then sort each attribute chronologically based on the exposure number. 
    We devise a baseline \textbf{PopUser}, in which the top user attributes are selected as explanations. Analogously, we build \textbf{PopItem} that produces the top item attributes for the explanation. 
    \item \textbf{FairKGAT}: uses FairKG4Rec~\cite{fu2020fairness} to mitigate the unfairness of explanations for a knowledge graph-enhanced recommender KGAT~\cite{wang2019kgat}.
    FairKG4Rec~\cite{fu2020fairness} is a generalized fairness-aware algorithm that controls the unfairness of explanation diversity in the recommendation model.
    KGAT~\cite{wang2019kgat} is a state-of-the-art knowledge graph-enhanced recommendation model that gives the best fairness performance in the original FairKG4Rec paper.
    \item \textbf{CEF}~\cite{DBLP:conf/sigir/GeTZXL0FGLZ22}: is the first work that explains fairness in recommendation. 
    It generates feature-based explanations for item exposure unfairness by perturbing user and item features and searches for features that change the fairness disparity. 
\end{itemize}
Note that to the best of our knowledge, \textbf{FairKGAT}~\cite{fu2020fairness} and \textbf{CEF}~\cite{DBLP:conf/sigir/GeTZXL0FGLZ22} are the only two existing methods designed for explainable fairness recommendation tasks.  

\subsubsection{Explanation Faithfulness Evaluation}\label{sec:evaluation}
We adopt the widely used erasure-based evaluation criterion~\cite{ge2021counterfactual} in Explainable AI to evaluate the explanation faithfulness.
The erasure-based evaluation identifies the contributions of explanations by measuring model performance changes after these explanations are removed.
As a result, one can tell whether the model actually relied on these particular explanations to make a prediction, i.e., faithful to the model.
In our experiments, we use the erasure-based evaluation to test (I) the recommendation performance change and (II) the recommendation fairness change after a set of attributes from the generated explanation is removed.
By doing so, we can identify whether our explanations are faithful to recommendation performance and fairness disparity.

Following~\cite{ge2021counterfactual}, we remove certain attributes from the generated explanations and evaluate the resulting recommendation performance. 
Therefore, in the starting evaluation point, we consider all attributes and add them to the user and item embeddings. 
We then remove certain attributes from the generated explanations to observe recommendation and fairness changes at later evaluation points.
In particular,
we first use historical user-item interactions to train a recommendation model through Eq.~\eqref{eq:recommendation model} to generate user and item embeddings.
Then, we fuse all attribute embeddings from Eq.~\eqref{eq:each layer} with the trained user and item embeddings.
The user and item embeddings after fusion are used to generate recommendation results at the starting evaluation point.
Thereafter, we conduct counterfactual reasoning using our \emph{CFairER} to generate attribute-level counterfactual explanations for model fairness. 
Those generated explanations are defined as the erasure set of attributes for each user/item.
Finally, we exclude the erasure set from attribute space, and fuse the embeddings of attributes after erasure with the trained user and item embeddings to generate new recommendation results. 

Given the recommendation results at each evaluation point, we use Normalized Discounted Cumulative Gain (NDCG)@$K$ and Hit Ratio (HR)@$K$ to measure the recommendation performance
As this work focuses on item exposure fairness in recommendations, we use two wildly-adopted item-side evaluation metrics, i.e., Head-tailed Rate (HT)@$K$ and Gini@$K$, for fairness evaluation.
HT@$K$ refers to the ratio of the head-tailed item number to the list length $K$.
Later HT@$K$ indicates that the model suffers from a more severe item exposure disparity by favoring items from the head-tailed (i.e., popular) group.
Gini@$K$ measures inequality within subgroups among the Top-$K$ recommendation list. 
Larger Gini@$K$ indicates the recommendation results are of higher inequality between the head-tailed and the long-tailed group.

\subsubsection{Implementation Details}\label{sec:implementation}

To demonstrate our \emph{CFairER}, we employ a simple matrix factorization (MF) as our recommendation model.
We train the MF using train/test/validate sets split from user-item interactions in datasets with 60\%/20\%/20\%. 
We optimize the MF using stochastic gradient descent (SGD)~\cite{bottou2012stochastic}. 
The same data splitting and gradient descent methods are applied in all baselines when required. 
Our graph representation module employs two graph convolutional layers with $\{64, 128\}$ output dimensions.
FairKGAT baseline also keep $2$ layers.
The graph representation module outputs embeddings for all user and item attributes with the embedding size $d=128$.
The embedding size for FairKGAT and CEF is also fixed as $d=128$.
The number of latent factors (as in Eq.~\eqref{eq:uv}) of MF is set equal to the embedding size of our graph representation module. 
To generate the starting evaluation point of erasure-based evaluation, we fuse attribute embeddings with the trained user and item latent factors based on element-wise product fusion.
The fused user and item embeddings are then used to produce Top-$K$ recommendation lists.

We train our counterfactual fairness explanation model with SGD based on the REINFORCE~\cite{sutton1999policy} policy gradient.
For baseline model compatibility, as CEF~\cite{DBLP:conf/sigir/GeTZXL0FGLZ22} requires pre-defined user-feature attention matrix and item-feature quality matrix, we follow previous work~\cite{wang2022reinforced} to regulate user/item attributes as user/item aspects and resort to analysis toolkit ``Sentires''~\footnote{https://github.com/evison/Sentires} to build the two matrices.
The hyper-parameters of our \emph{CFairER} and all baselines are chosen by the grid search, including learning rate, $L_2$ norm regularization, discount factor $\gamma$, etc.
The disparity change threshold $\epsilon$ in Eq.~\eqref{eq:reward} of our \emph{CFairER} is determined by performing a grid search on the validation set. 
This enables us to choose the optimal value for a variety of recommendation tasks, including but not limited to business (\texttt{Yelp} dataset), movie (\texttt{Douban Movie} dataset), and music (\texttt{Last-FM} dataset) recommendations.
After all models have been trained, we freeze the model parameters and generate explanations accordingly.
We report the erasure-based evaluation results by recursively erasing top $E$ attributes from the generated explanations.
The erasure length $E$ is chosen from $E=[5, 10, 15, 20]$.
The recommendation and fairness performance of our \emph{CFairER} and baselines under different $E$ is reported in Table~\ref{tab:overall}.

\subsection{Explanation Faithfulness (RQ1, RQ2)}

\begin{table*}[htbp]
\caption{Recommendation and fairness performance after erasing top $E=[5,10,20]$ attributes from explanations. 
$\uparrow$ represents larger values are desired for better performance, while $\downarrow$ indicates smaller values are better. 
Bold and underlined numbers are the best results and the second-best results, respectively.
}\label{tab:overall}
\centering
\setlength{\tabcolsep}{5pt}
\resizebox{\textwidth}{!}{
\begin{tabular}
    {c ccc ccc ccc ccc} \toprule
    \multirow{2}{*}{Method} 
    & \multicolumn{3}{c}{NDCG@40 $\uparrow$} 
    & \multicolumn{3}{c}{Hit Ratio (HR)@40 $\uparrow$} 
    & \multicolumn{3}{c}{Head-tailed Rate (HT)@40 $\downarrow$}
    & \multicolumn{3}{c}{Gini@40 $\downarrow$}\\\cmidrule(lr){2-4} \cmidrule(lr){5-7} \cmidrule(lr){8-10} \cmidrule(lr){11-13}
 & $E=5$ & $E=10$ & $E=20$ & $E=5$ & $E=10$ & $E=20$ & $E=5$ & $E=10$ & $E=20$ & $E=5$ & $E=10$ & $E=20$ \\\midrule 

\multicolumn{13}{c}{\texttt{Yelp}} \\\midrule
\textbf{RDExp} &0.0139  & 0.0125  &0.0118  &0.1153  & 0.1036  &0.1029  &0.1994  & 0.1976   &0.1872  &0.3870 &0.3894   &0.3701 \\
\textbf{PopUser}  &0.0141  & 0.0136  &0.0128  &0.1183  &0.1072  &0.1067  &0.1776  &0.1767 &0.1718  &0.3671  & 0.3642  &0.3495 \\
\textbf{PopItem}  &0.0147 & 0.0139  &0.0131  &0.1182  & 0.1093 & 0.1084 &0.1793  & 0.1846 &0.1848  &0.3384  &0.3370   &0.3359 \\
\textbf{FairKGAT}  & 0.0153  & 0.0141  & 0.0138  &0.1384  &0.1290  &0.1277  &0.1802  &0.1838  &0.1823  &0.3671  &0.3508  &0.3542 \\
\textbf{CEF}  & \underline{0.0254}  & \underline{0.0247}  &\underline{0.0231}  &\underline{0.1572}  &\underline{0.1608}  &\underline{0.1501}  &\underline{0.1496}  & \underline{0.1455}  &\underline{0.1420}  &\underline{0.3207}  &\underline{0.3159}  &\underline{0.3088} \\
\textbf{CFairER}  & \textbf{0.0316} & \textbf{0.0293} & \textbf{0.0291} & \textbf{0.1987}  & \textbf{0.1872} &\textbf{0.1868 } &\textbf{0.1345}  &\textbf{0.1322}  &\textbf{0.1301}  &\textbf{0.2366}   &\textbf{0.2068}  &\textbf{0.1974} \\
\cline{2-13}
\midrule

\multicolumn{13}{c}{\texttt{Douban Movie}} \\\midrule
\textbf{RDExp} &0.0390  & 0.0346  &0.0351  &0.1278  & 0.1170  &0.1172  & 0.1932 &0.1701   &0.1693  &0.3964  & 0.3862  &0.3741 \\
\textbf{PopUser}  &0.0451  &0.0352  &0.0348  &0.1482  & 0.1183  &0.1174  &0.1790  &0.1674  &0.1658  &0.3684  &0.3591  &0.3562 \\
\textbf{PopItem}  &0.0458  &0.0387  &0.0379  &0.1523  & 0.1219 & 0.1208  &0.1831  &0.1458   &0.1383  &0.3664  &0.3768  &0.3692 \\
\textbf{FairKGAT}  &0.0534  &0.0477  &0.0421  & 0.1602 &0.1377  &0.1308  &0.1654  &0.1573  &0.1436  &0.3590  &0.3472  &0.3483 \\
\textbf{CEF}  &\underline{0.0831}  &\underline{0.0795}  &\underline{0.0809}  &\underline{0.1949}  &\underline{0.1973}  &\underline{0.1901}  &\underline{0.1043}  &\textbf{0.0998}  &\textbf{0.0945}  &\underline{0.3079}  &\underline{0.2908}  &\underline{0.3001} \\
\textbf{CFairER}  &\textbf{0.1290}  & \textbf{0.0921}  &\textbf{0.0901}  &\textbf{0.2706}  & \textbf{0.2441}  &\textbf{0.2238 } &\textbf{0.0841}   &\underline{0.1183}    &\underline{0.1101} &\textbf{0.2878}  &\textbf{0.2648}  &\textbf{0.2593} \\
\cline{2-13}
\midrule

\multicolumn{13}{c}{\texttt{Last-FM}} \\\midrule
\textbf{RDExp} &0.0857  &0.0568  &0.0592  & 0.8436  &0.7915  &0.7831  &0.7884  & 0.7732  &0.7691  &0.3707  &0.3531  &0.3540 \\
\textbf{PopUser}  &0.0786  &0.0432  &0.0431  &0.7787  &0.7697  &0.7604  &0.7979  &0.8064   &0.7942  &0.3862  &0.3729  &0.3761 \\
\textbf{PopItem}  &0.0792  & 0.0479 &0.0435  &0.7803  &0.7961  &0.7914  &0.7689  &0.7638  &0.7573  &0.3673  &0.3602 &0.3618\\
\textbf{FairKGAT}  &0.0832  & 0.0594 &0.0621  &0.8063    &0.7938  &0.8165  &0.7451  &0.7342  &0.7408  &0.3580  &0.3458 &0.3433\\
\textbf{CEF}  &\underline{0.0962}  &\underline{0.1037}  &\underline{0.1001}  &\underline{0.8592}  &\underline{0.8408}  &\underline{0.8509}  &\underline{0.6873}  &\underline{0.6092}  &\textbf{0.5601}  &\underline{0.3308}  &\underline{0.3298}  &\underline{0.3375} \\
\textbf{CFairER}  &\textbf{0.1333}  &\textbf{0.1193}  &\textbf{0.1187}  &\textbf{0.9176}  &\textbf{0.8867}  &\textbf{0.8921}  &\textbf{0.6142}  &\textbf{0.5865}  &\underline{0.5737}  &\textbf{0.2408}  &\textbf{0.2371}  &\textbf{0.2385} \\
\cline{2-13}
\bottomrule
\end{tabular}
}
\end{table*}

\begin{figure*}[!h]
\centering
\begin{minipage}[t]{0.45\textwidth}
\centering
\includegraphics[width=\textwidth]{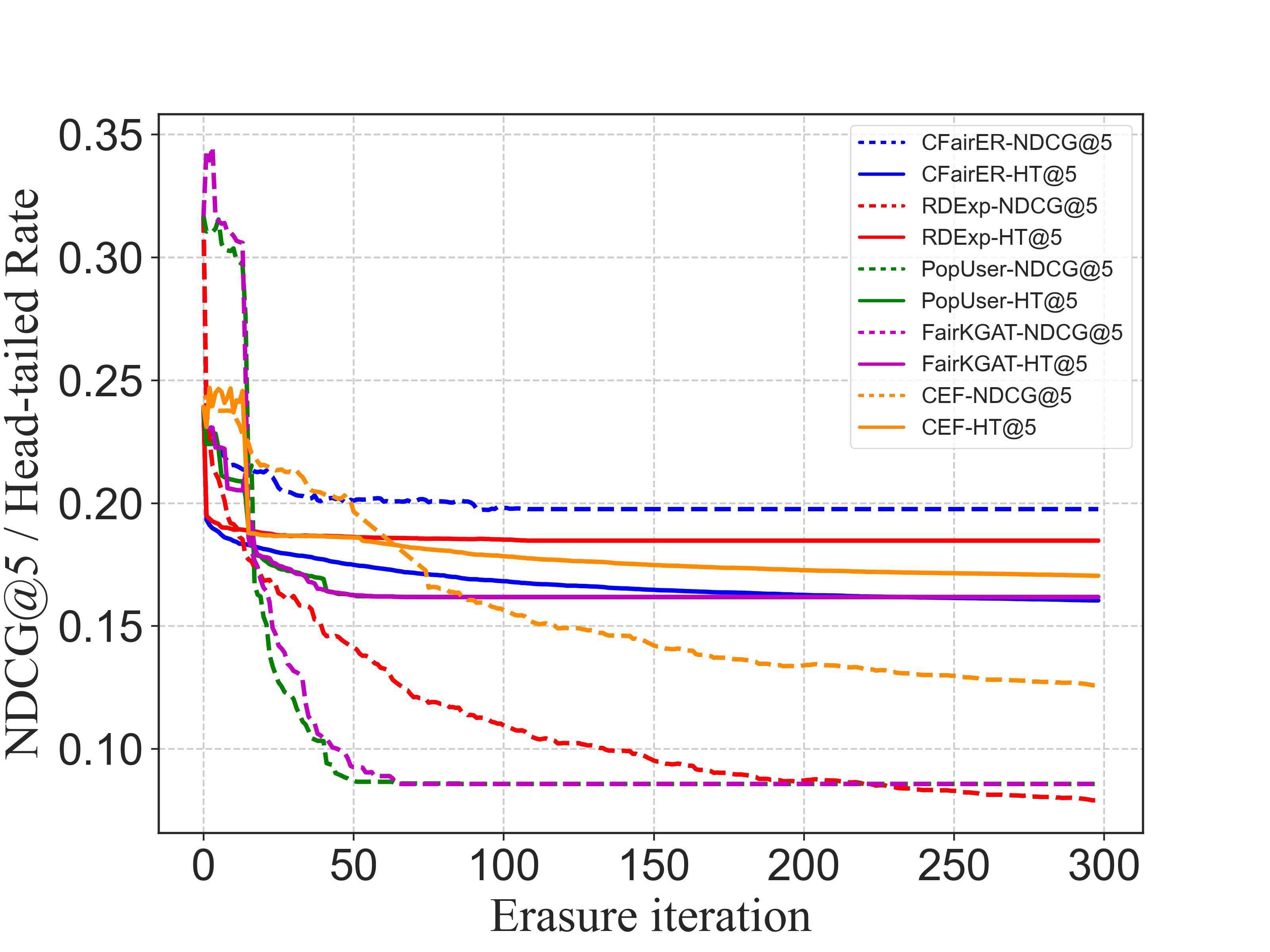}
\subcaption{\centering NDCG@5 and Head-tailed Rate@5 on \texttt{Yelp}}
\end{minipage}
\begin{minipage}[t]{0.45\textwidth}
\centering
\includegraphics[width=\textwidth]{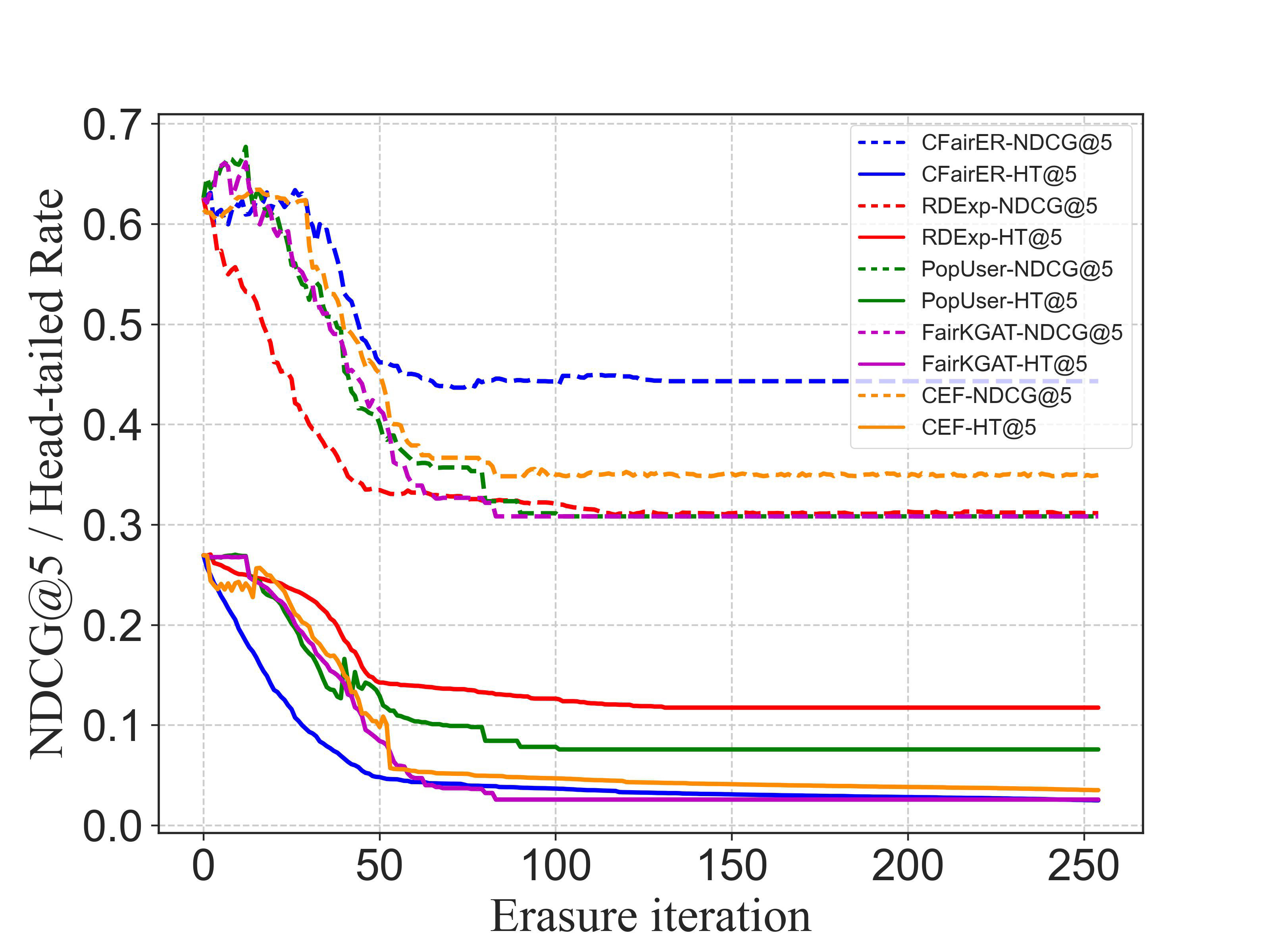}
\subcaption{\centering NDCG@5 and Head-tailed Rate@5 on \texttt{Douban Movie}}
\end{minipage}
\begin{minipage}[t]{0.45\textwidth}
\centering
\includegraphics[width=\textwidth]{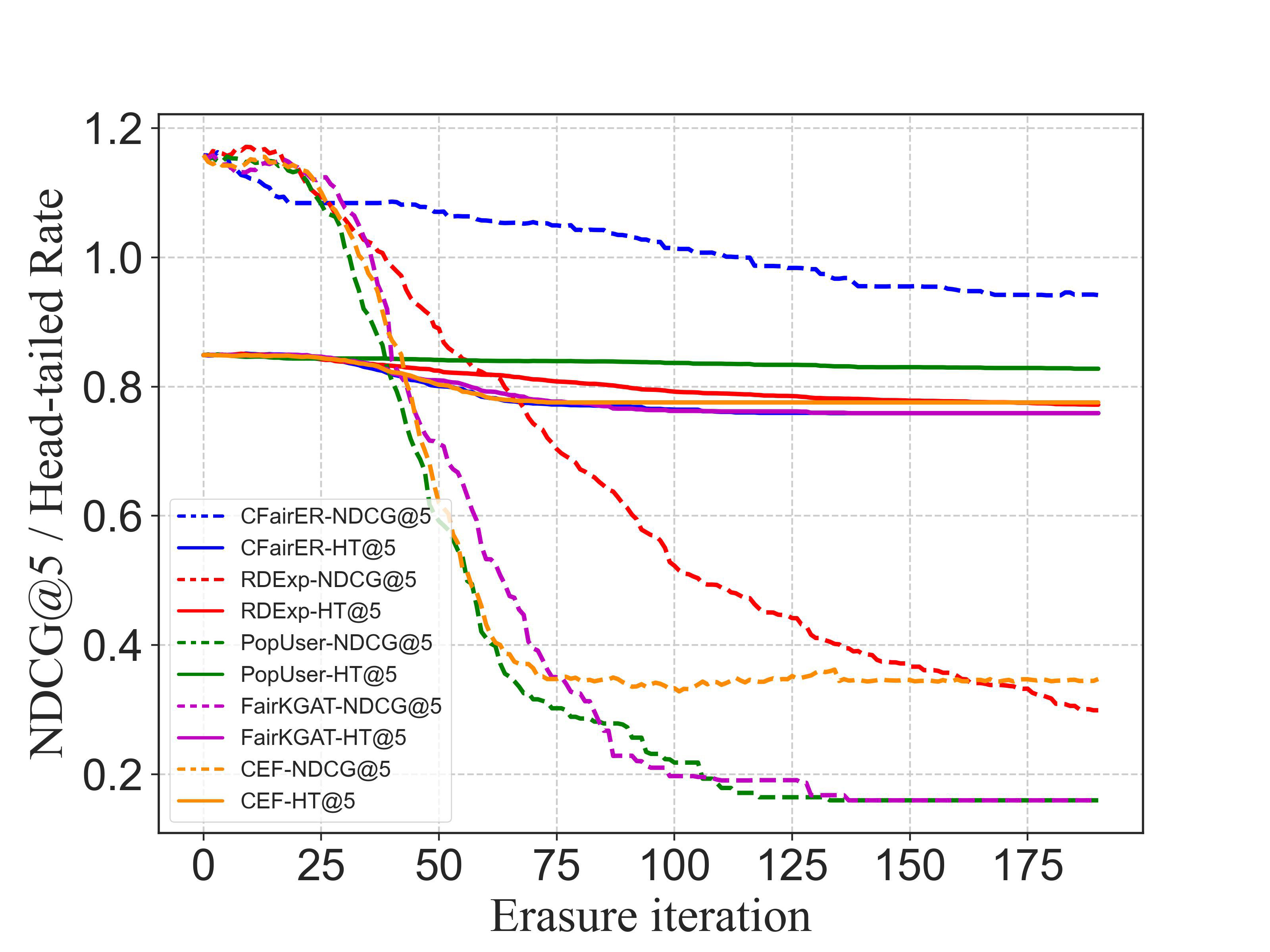}
\subcaption{\centering NDCG@5 and Head-tailed Rate@5 on \texttt{Last-FM}}
\end{minipage}
\begin{minipage}[t]{0.45\textwidth}
\centering
\includegraphics[width=\textwidth]{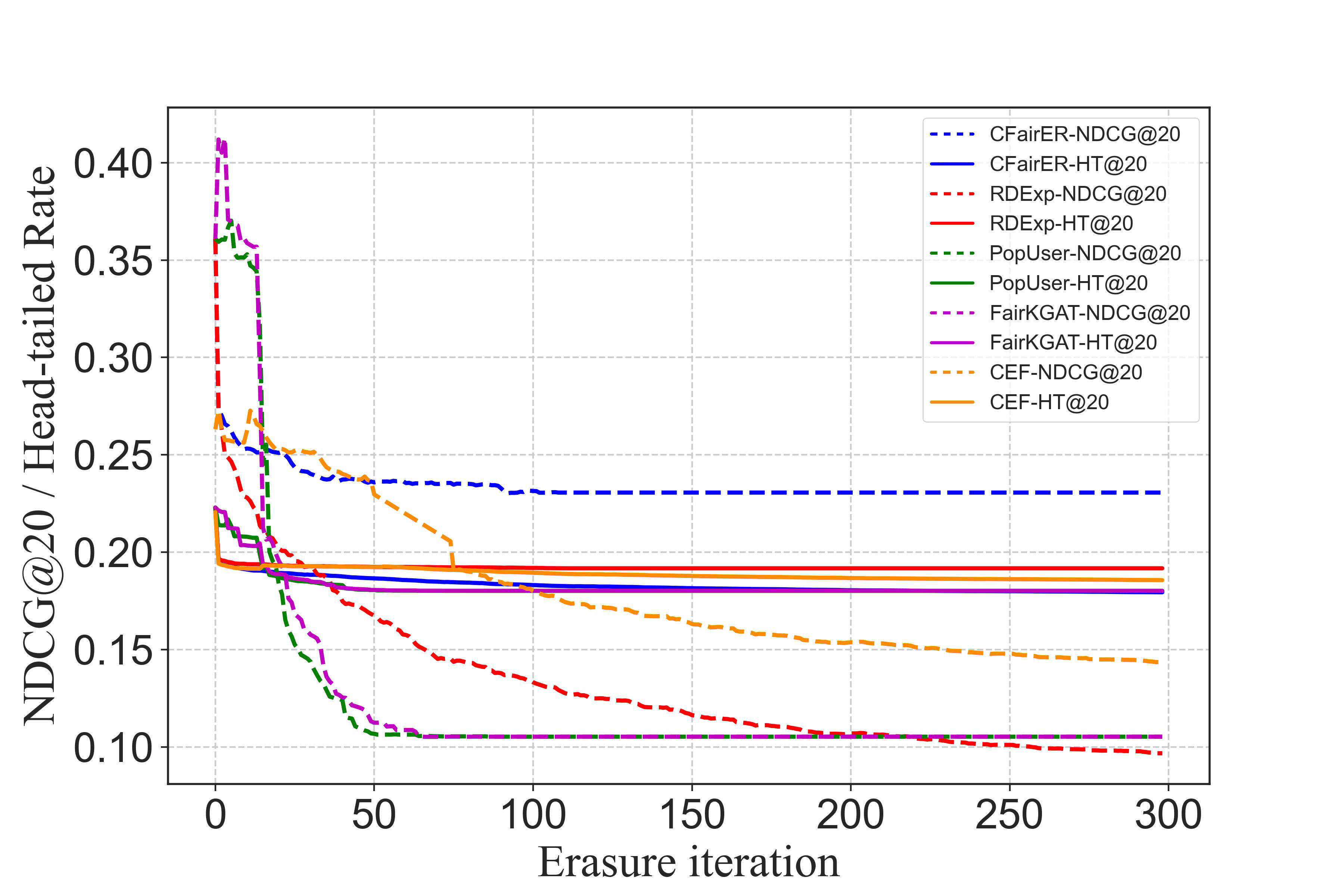}
\subcaption{\centering NDCG@20 and Head-tailed Rate@20 on \texttt{Yelp}}
\end{minipage}
\begin{minipage}[t]{0.45\textwidth}
\centering
\includegraphics[width=\textwidth]{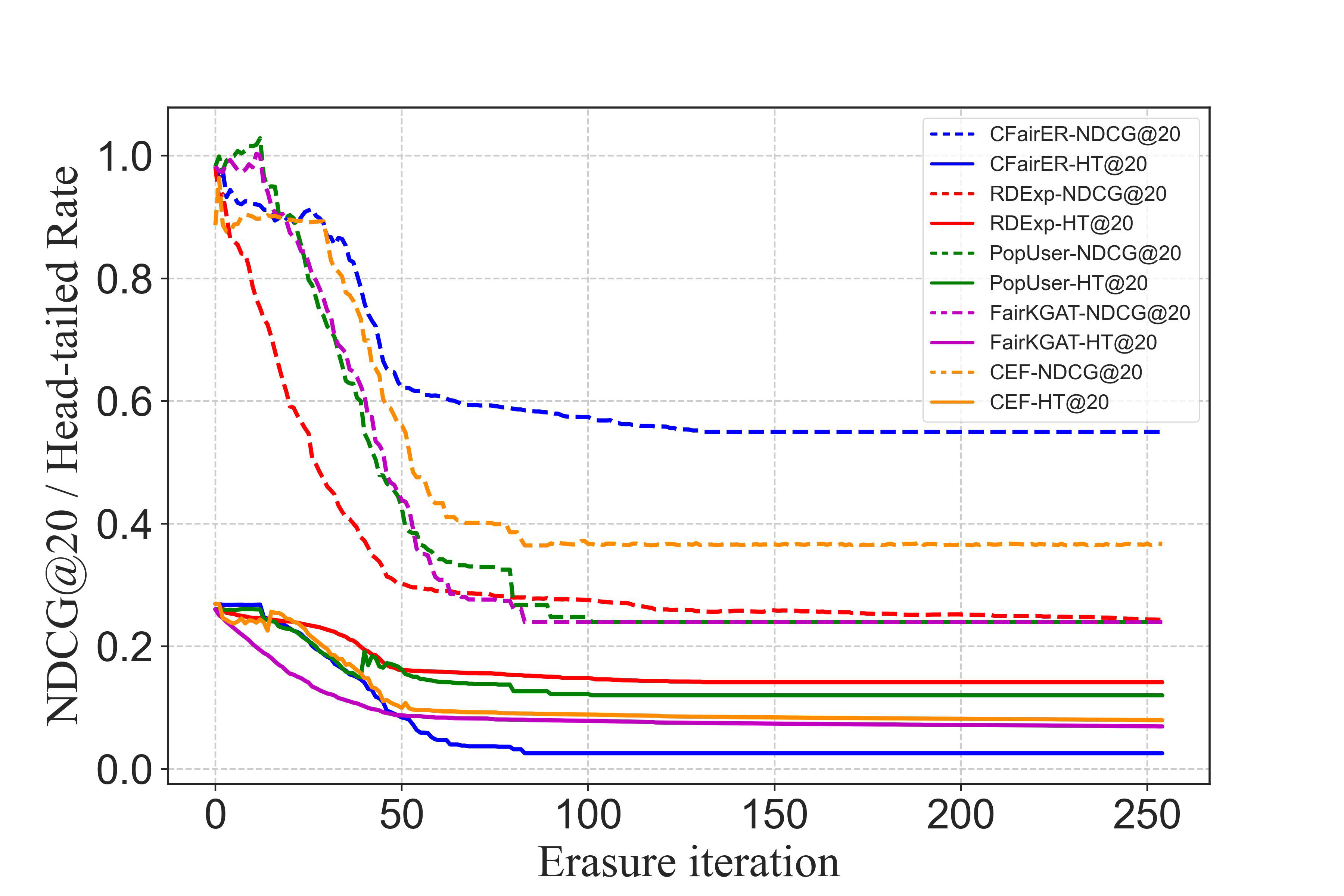}
\subcaption{\centering NDCG@20 and Head-tailed Rate@20 on \texttt{Douban Movie}}
\end{minipage}
\begin{minipage}[t]{0.45\textwidth}
\centering
\includegraphics[width=\textwidth]{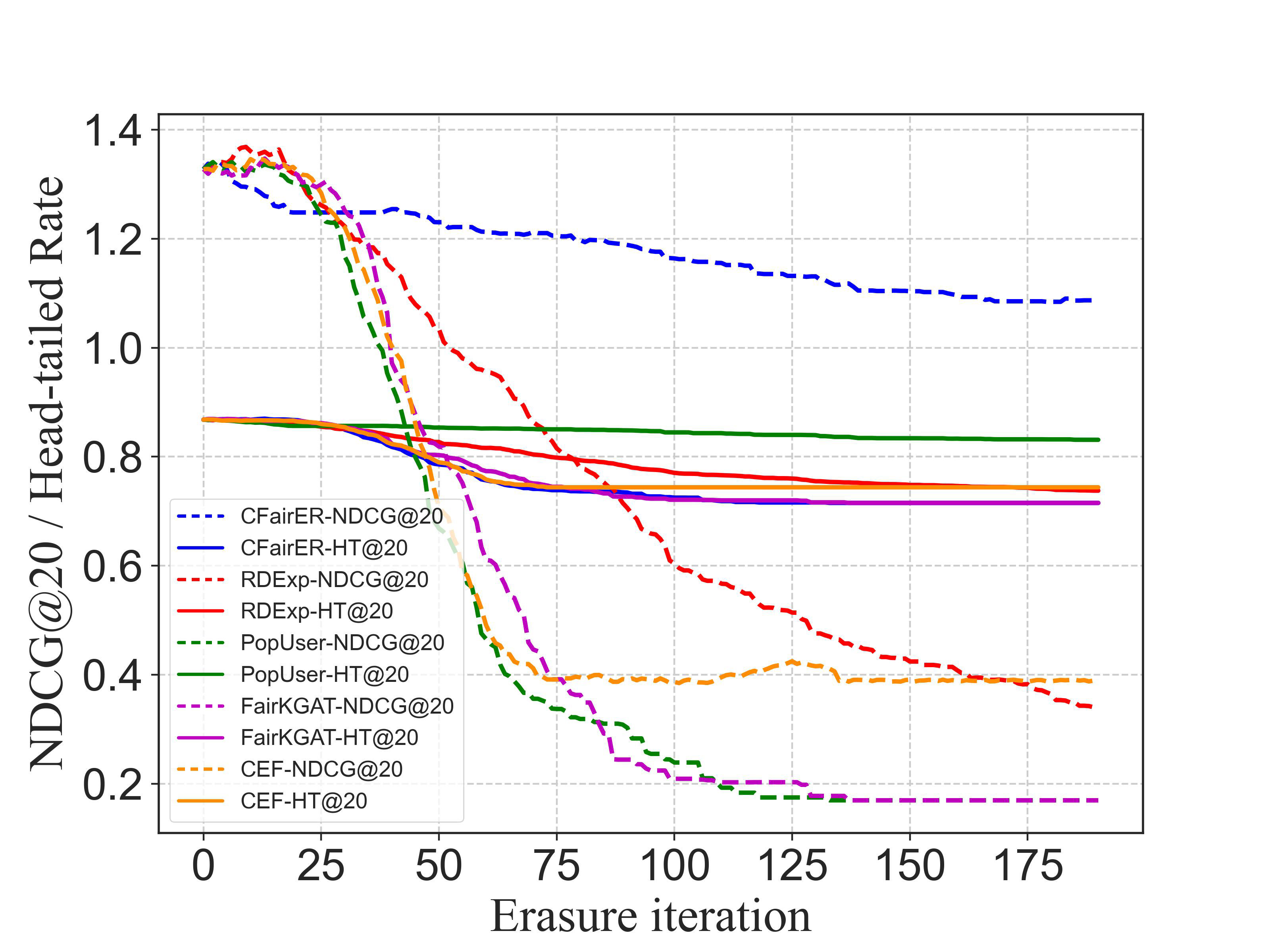}
\subcaption{\centering NDCG@20 and Head-tailed Rate@20 on \texttt{Last-FM}}
\end{minipage}
\caption{Erasure-based evaluation on Top-5 and Top-20 recommendations. 
NDCG@$K$ values reveal the recommendation performance of models, while Head-tailed Rate@$K$ values reflect model fairness. 
NDCG@$K$ values are multiplied with $10$ for better presentation.
Each data point is generated while cumulatively erasing the top $10$ (i.e., $E=10$) attributes in explanations.}
\label{fig:overall}
\end{figure*}

We plot fairness and recommendation performance changes of our \emph{CFairER} and baselines while erasing attributes from explanations in Figure~\ref{fig:overall}.
Each data point in Figure~\ref{fig:overall} is generated by cumulatively erasing a batch of attributes.
Those erased attributes are selected from the top 10 (i.e., $E=10$) attribute sets of the explanation lists provided by each method.\footnote{For example, given $n$ explanation lists, the number of erasure attributes is $n \times 10$. We cumulatively erase $m$ attributes in one batch within in total $(n \times 10) / m $ iterations.}
As PopUser and PopItem baselines enjoy very similar data trends, we choose not to present them simultaneously in Figure~\ref{fig:overall}.
Table~\ref{tab:overall} presents recommendation and fairness performance after erasing $E = [5, 10, 20]$ attributes in explanations.
Larger NDCG@$K$ and Hit Ratio @$K$ values indicate better recommendation performance while smaller Head-tailed Rate@$K$ and Gini@$K$ values represent better fairness. 
Analyzing Figure~\ref{fig:overall} and Table~\ref{tab:overall}, we have the following findings.

Amongst all methods, our \emph{CFairER} achieves the best recommendation and fairness performance after erasing attributes from our explanations on all datasets.
For instance, \emph{CFairER} beats the strongest baseline CEF by 25.9\%, 24.4\%, 8.3\% and 36.0\% for NDCG@40, Hit Ratio@40, Head-tailed Rate@40 and Gini@40 with erasure length $E=20$ on \texttt{Yelp}.
This indicates that explanations generated by \emph{CFairER} are faithful to explaining unfair factors while not harming recommendation accuracy.
Unlike CEF and FairKGAT, which generate explanations based on perturbing input features and adding fair-related constraints, \emph{CFairER} generates counterfactual explanations by inferring minimal attributes contributing to fairness changes. 
As a counterfactual explanation is minimal, it only discovers attributes that well-explain the model fairness while filtering out tedious ones that affect the recommendation accuracy.
Another interesting finding is that
PopUser and PopItem perform even worse than RDExp (i.e., randomly selecting attributes) on \texttt{Last-FM}.
This is because recommending items with popular attributes would deprive the exposure of less-noticeable items, causing serious model unfairness and degraded recommendation performance.

In general, the fairness of all models consistently improves while erasing attributes from explanations, shown by the decreasing trend of Head-tailed Rate@$K$ values in Figure~\ref{fig:overall}.
This is because erasing attributes will alleviate the discrimination against users and items from disadvantaged groups (e.g., gender group, brand group), making more under-represented items to be recommended. 
Unfortunately, 
we can also observe the downgraded recommendation performance of all models in both Figure~\ref{fig:overall} and Table~\ref{tab:overall}.
For example, in Figure~\ref{fig:overall}, the NDCG@5 of CEF drops from approximately 1.17 to 0.60 on \texttt{Last-FM} at erasure iteration 0 and 50.
This is due to the well-known fairness-accuracy trade-off issue, in which the fairness constraint could be achieved with a sacrifice of recommendation performance.
Facing this issue, both baselines suffer from huge declines in recommendation performance, as in Table~\ref{tab:overall}. 
On the contrary, our \emph{CFairER} still enjoys favorable recommendation performance and outperforms all baselines.
Besides, the decline rates of our \emph{CFairER} are much slower than baselines on both datasets in Figure~\ref{fig:overall}.
We hence conclude that the attribute-level explanations provided by our \emph{CFairER} can achieve a much better fairness-accuracy trade-off than other methods. 
This is because our \emph{CFairER} uses counterfactual reasoning to generate minimal but vital attributes as explanations for model fairness.
Those attributes produced by \emph{CFairER} are true reasons for unfairness but not the ones that affect the recommendation accuracy. 

\subsection{Ablation and Parameter Analysis (RQ3)}

\begin{table}[htbp]
\caption{Ablation Study on \emph{CFairER}. 
Erasure length $E$ is fixed as $E=20$.
$\neg$ represents the corresponding module is removed.
$A \to B$ represents $A$ is replaced by $B$.
$\pm$ indicates the increase or decrease percentage of the variant compared with \emph{CFairER}. 
}\label{tb:ablation}
 \resizebox{\textwidth}{!}{
\begin{tabular}{c c c c c}
\hline
Variants & NDCG@20$\uparrow$ & HR@20 $\uparrow$ & HT@20$\downarrow$ & Gini@20$\downarrow$
\\ \hline
\multicolumn{5}{c}{\texttt{Yelp}} \\\midrule
CFairER & 0.0238  & 0.1871  & 0.1684  & 0.1990  \\  
 CFairER $\neg$ Attentive Action Pruning &0.0164($-31.1\%$)  &0.1682($-10.1\%$)  &0.1903($+13.0\%$)  &0.2159($+8.5\%$)  \\  
 CRM loss $\to$ Cross-entropy~\cite{zhang2018generalized} loss &0.0197($-17.2\%$) &0.1704($-8.9\%$)  &0.1841($+9.3\%$)  &0.2101($+5.6\%$)  \\ 
\hline

\multicolumn{5}{c}{\texttt{Douban Movie}} \\\midrule
CFairER & 0.0583  & 0.2043  & 0.1149  & 0.2871  \\  
 CFairER $\neg$ Attentive Action Pruning &0.0374($-35.9\%$)  &0.1537($-24.8\%$)  &0.1592($+38.6\%$)  &0.3574($+24.5\%$)  \\  
 CRM loss $\to$ Cross-entropy~\cite{zhang2018generalized} loss &0.0473($-18.9\%$) &0.1582($-22.6\%$)  &0.1297($+12.9\%$)  &0.3042($+6.0\%$)  \\   
\hline
\multicolumn{5}{c}{\texttt{Last-FM}} \\\midrule
CFairER & 0.1142  & 0.7801  & 0.6914  & 0.2670  \\  
 CFairER $\neg$ Attentive Action Pruning &0.0987($-13.6\%$)  &0.7451($-4.5\%$)  &0.7833($+13.3\%$)  &0.2942($+10.2\%$)  \\  
 CRM loss $\to$ Cross-entropy~\cite{zhang2018generalized} loss &0.0996($-12.8\%$) &0.7483($-4.1\%$)  &0.7701($+11.4\%$)  &0.2831($+6.0\%$)  \\ 
\hline
\end{tabular}}
\end{table}

We first conduct an in-depth ablation study on the ability of our \emph{CFairER} to achieve sample efficiency and bias alleviation.
Our \emph{CFairER} includes two contributing components,
namely, attentive action pruning (cf. Section~\ref{sec:CEF}) and counterfactual risk minimization-based optimization (cf. Section~\ref{sec:optimization}).
We evaluate our \emph{CFairER} with different variant combinations and show our main findings below.

\subsubsection{Sample Efficiency of Attentive Action Pruning}
 
Our attentive action pruning reduces the action search space by specifying varying importance of attributes for each state.
As a result, the sample efficiency can be increased by filtering out irrelevant attributes to promote an efficient action search.
To demonstrate our attentive action pruning, we test \emph{CFairER} without ($\neg$) the attentive action pruning (i.e., \emph{CFairER $\neg$ Attentive Action Pruning}), in which the candidate actions set absorbs all attributes connected with the current user and items. 
Through Table~\ref{tb:ablation}, we observed that removing the attentive action pruning downgrades \emph{CFairER} performance, which validates the superiority of our attentive action pruning in improving fair recommendations.
This is because attentive action pruning filters out irrelevant items based on their contributions to the current state, resulting in enhanced sample efficiency.
Moreover, the performance of \emph{CFairER} after removing the attentive action pruning downgrades severely on \texttt{Douban Movie}. 
This is because \texttt{Douban Movie} has the largest number of attributes compared with the other two datasets (cf. Table~\ref{tb:dataset}), which challenges our \emph{CFairER} to find suitable attributes as fairness explanations.
These findings suggest the superiority of applying attentive action pruning in fairness explanation learning, especially when the attribute size is large.

\subsubsection{Bias Alleviation with Counterfactual Risk Minimization}

Our \emph{CFairER} is optimized with a counterfactual risk minimization (CRM) loss to achieve unbiased policy optimization.
The CRM loss (cf. Eq.~\eqref{eq:crm}) corrects the discrepancy between the explanation policy and logging policy, thus alleviating the policy distribution bias in the off-policy learning setting.
To demonstrate the CRM loss, 
we apply our \emph{CFairER} with cross-entropy (CE)~\cite{zhang2018generalized} loss (i.e., \emph{CRM loss $\to$ Cross-entropy loss}) to show how it performs compared with \emph{CFairER} on the CRM loss. 
We observe our \emph{CFairER} with CRM loss consistently outperforms the counterpart with CE loss on both fairness and recommendation performance.
The sub-optimal performance of our \emph{CFairER} with CE loss indicates that the bias issue in the off-policy learning can lead to downgraded performance for the learning agent. 
On the contrary, our \emph{CFairER} takes advantage of CRM to learn a high-quality explanation policy.
We hence conclude that performing unbiased optimization with CRM is critical to achieving favorable fairness explanation learning.

\begin{figure}[htbp]
\centering
\begin{minipage}[t]{0.43\textwidth}
\centering
\includegraphics[width=\textwidth]{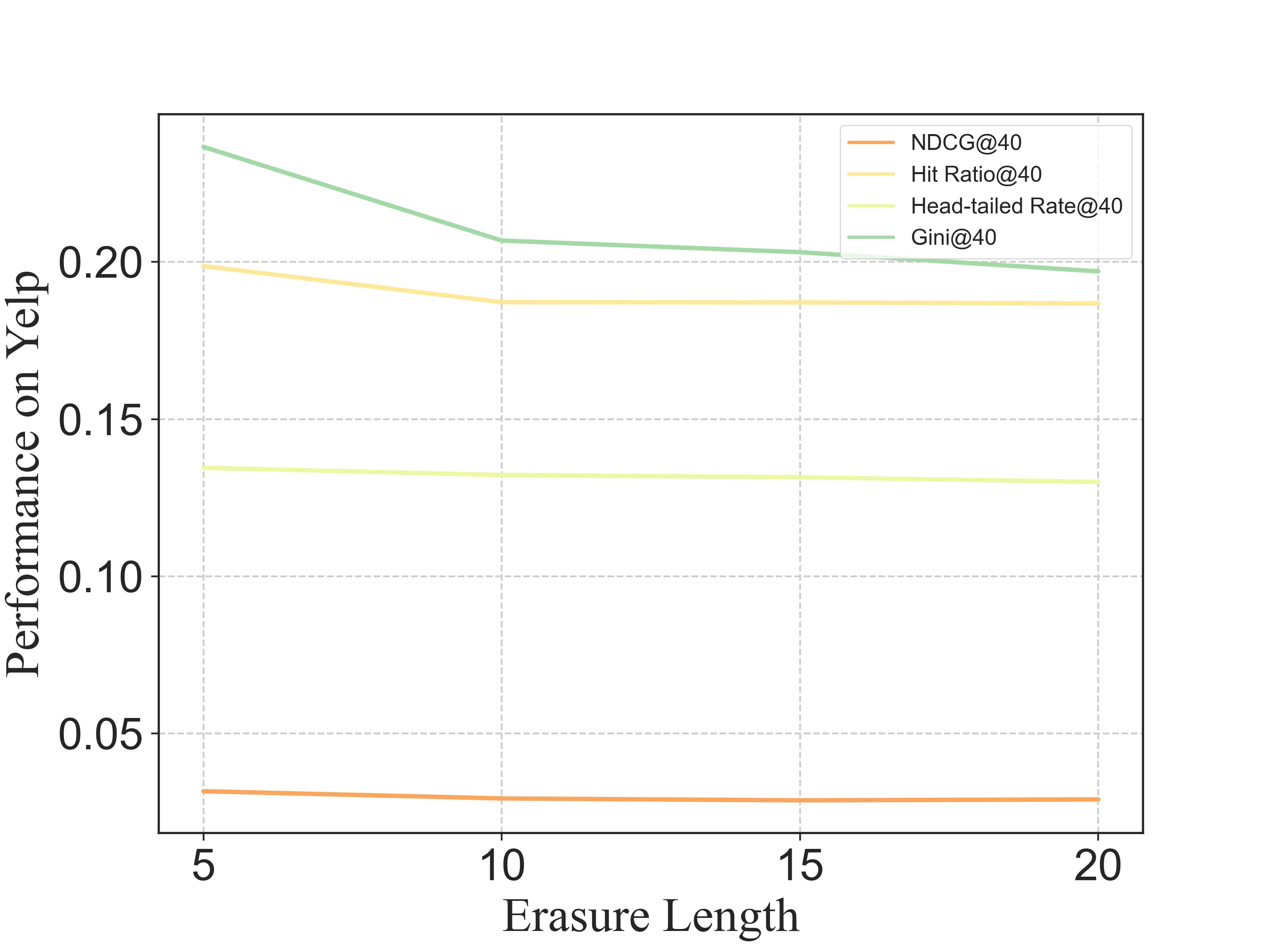}
\subcaption{\centering Impact of $E$ on \texttt{Yelp}.}
\end{minipage}
\begin{minipage}[t]{0.43\textwidth}
\centering
\includegraphics[width=\textwidth]{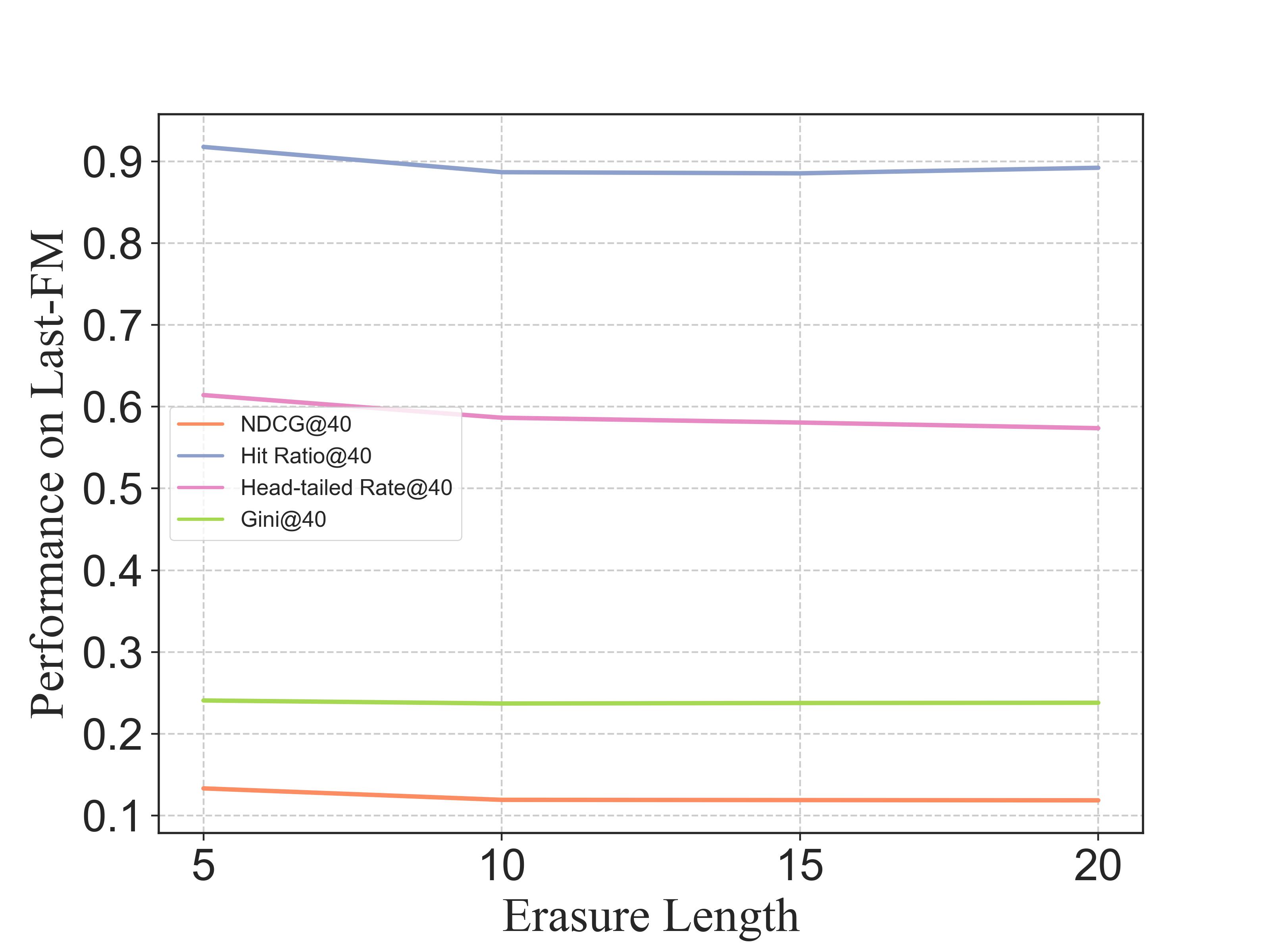}
\subcaption{ \centering Impact of $E$ on \texttt{Last-FM}.}
\end{minipage}

\begin{minipage}[t]{0.43\textwidth}
\centering
\includegraphics[width=\textwidth]{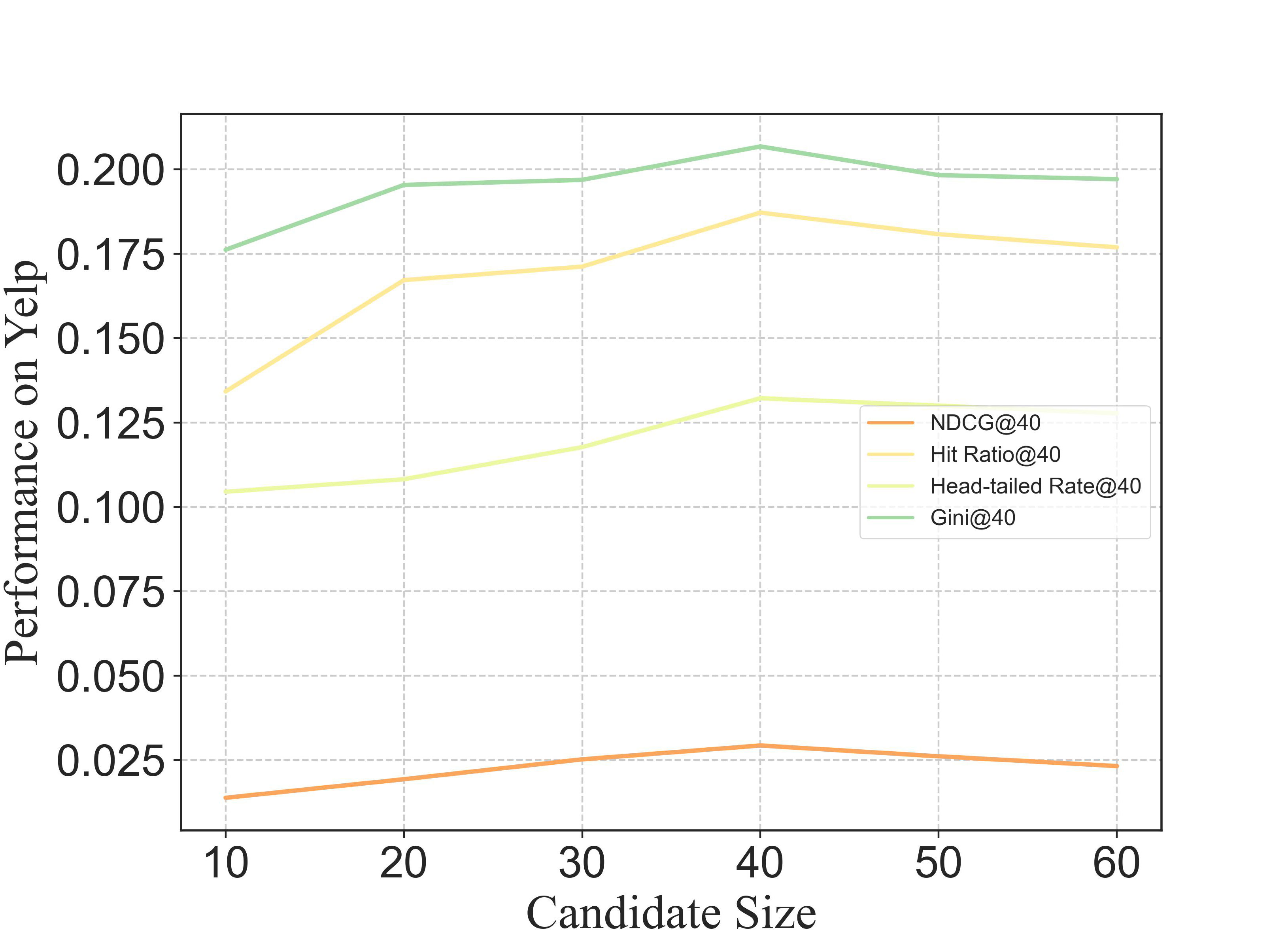}
\subcaption{ \centering Impact of $n$ on \texttt{Yelp}.}
\end{minipage}
\begin{minipage}[t]{0.43\textwidth}
\centering
\includegraphics[width=\textwidth]{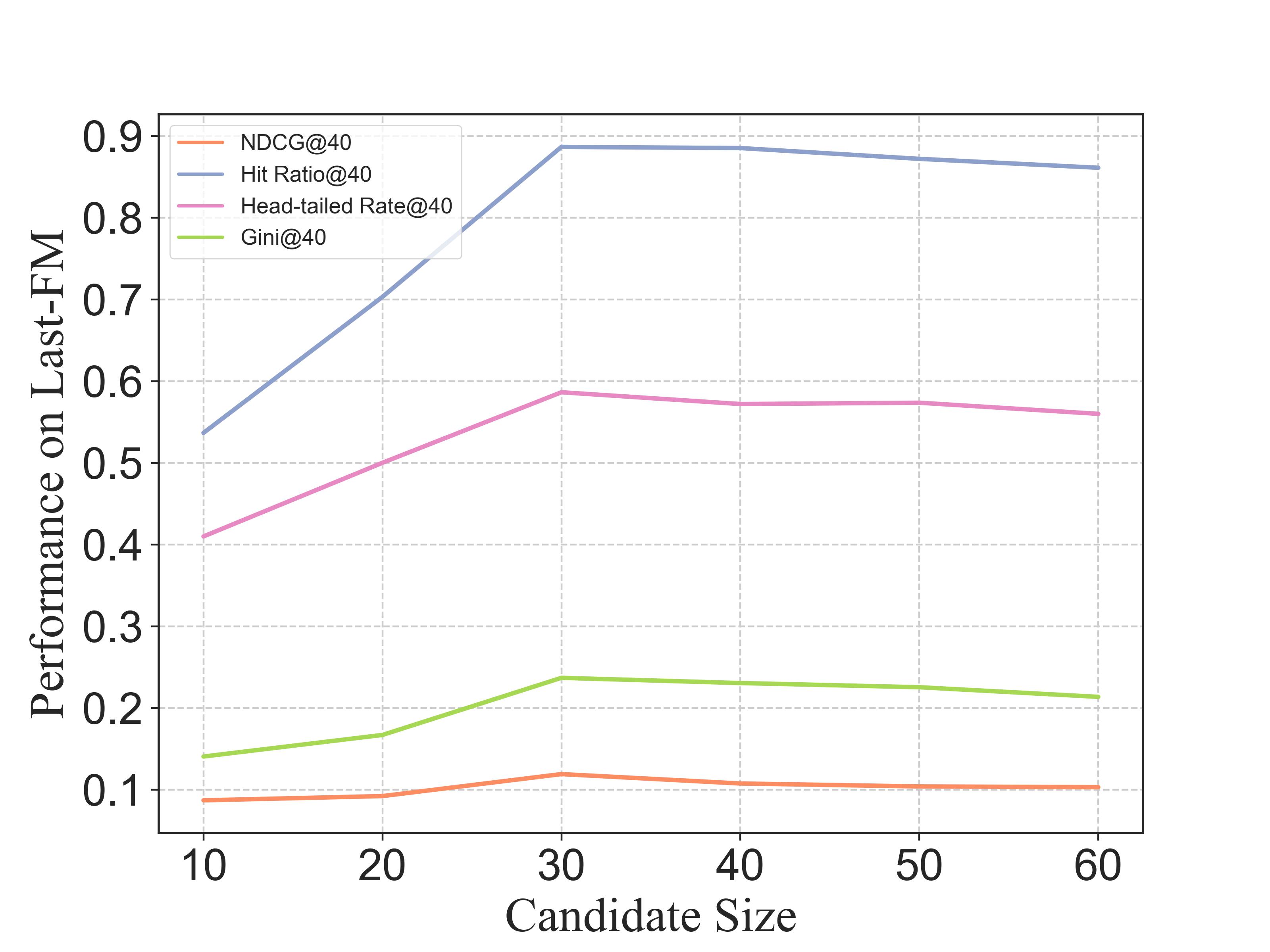}
\subcaption{ \centering Impact of $n$ on \texttt{Last-FM}.}
\end{minipage}
\caption{Impacts of parameters $E$ and $n$ on \emph{CFairER}.}
\label{fig:case-sens}
\end{figure}

\subsubsection{Parameter Analysis}
We also conduct a parameter analysis on how erasure length $E$ (cf. Section~\ref{sec:evaluation}) and candidate size $n$ (as in Eq.~\eqref{eq:candidate}) impact \emph{CFairER}. 
Figure~\ref{fig:case-sens} (a) and Figure~\ref{fig:case-sens} (b) report \emph{CFairER} performance w.r.t. $E=[5, 10, 15, 20]$.
Apparently, the performance of \emph{CFairER} demonstrates decreasing trends from $E=5$, then becomes stable after $E=10$.
The decreased performance is due to the increasing erasure of attributes found by our generated explanations. 
This indicates that our \emph{CFairER} can find valid attribute-level explanations that impact fair recommendations. 
The performance of \emph{CFairER} degrades slightly after the bottom, then becomes stable. 
This is reasonable since the attributes number provided in datasets are limited, while increasing the erasure length would allow more overlapping attributes with previous erasures to be found.

By varying candidate size $n$ from $n=[10, 20, 30, 40, 50, 60]$ in Figure~\ref{fig:case-sens} (c) (d), 
we observe that \emph{CFairER} performance first improves drastically as candidate size increases on both datasets.
The performance of our \emph{CFairER} reaches peaks at $n=40$ and $n=30$ on \texttt{Yelp} and \texttt{Last-FM}, respectively. 
After the peaks, we can witness a downgraded model performance by increasing the candidate size further.
We consider the poorer performance of \emph{CFairER} before reaching peaks is due to the limited candidate pool, i.e., insufficient attributes limit the exploration ability of \emph{CFairER} to find appropriate candidates as fairness explanations. 
Meanwhile, a too-large candidate pool (e.g., $n=60$) would offer more chances for the agent to select inadequate attributes as explanations.
Based on the two findings, we believe it is necessary for our \emph{CFairER} to carry the attentive action search, such as to select high-quality attributes as candidates based on their contributions to the current state. 

\subsubsection{Time Complexity and Computation Costs}
For time complexity, our recommendation model (cf. Section~\ref{sec:rec}) performs matrix factorization with a complexity of $O(|\mathcal{O}|)$. 
For the graph representation module (cf. Section~\ref{sec:graph representation}), establishing node representations has complexity $O(\sum_{l=1}^L (|\mathcal{G}|+|\mathcal{O}^{+}|) d_l d_{l-1})$. 
For the off-policy learning process (cf. Section~\ref{sec:CEF}), the complexity is mainly determined by the attention score calculation, which has a time complexity of $O(2T|\mathcal{O}^{+}| |\tilde{\mathcal{N}}_e| d^2)$. 
The total time complexity is $O(|\mathcal{O}|+ \sum_{l=1}^L(|\mathcal{G}|+|\mathcal{O}^{+}|) d_l d_{l-1}+2T|\mathcal{O}^{+}| n_2 d^2)$. 
We evaluated the running time of FairKGAT and CEF baselines on the large-scale \texttt{Yelp} dataset.
The corresponding results are 232s and 379s per epoch, respectively.
\emph{CFairER} has a comparable cost of 284s per epoch to these baselines. Considering that our \emph{CFairER} achieves superior explainability improvements compared to the baselines, we believe that the increased cost of, at most, 52s per epoch is a reasonable trade-off.

\section{Conclusion}
We propose \emph{CFairER}, a reinforcement learning-based fairness explanation learning framework over a HIN.
Our \emph{CFairER} generates counterfactual explanations as minimal sets of real-world attributes to explain item exposure fairness.
We design a counterfactual fairness explanation model to discover high-quality counterfactual explanations, driven by an attentive action pruning to reduce the search space and a counterfactual reward to enable counterfactual reasoning.
Extensive evaluations on three benchmark datasets demonstrate \emph{CFairER}’s ability to find faithful explanations for fairness and balance the fairness-accuracy trade-off.

\begin{acks}
This work is supported by the Australian Research Council (ARC) under Grant No. DP220103717, LE220100078, LP170100891 and DP200101374.
\end{acks}

\bibliographystyle{ACM-Reference-Format}
\bibliography{sample-base}


\begin{thebibliography}{62}


\ifx \showCODEN    \undefined \def \showCODEN     #1{\unskip}     \fi
\ifx \showDOI      \undefined \def \showDOI       #1{#1}\fi
\ifx \showISBNx    \undefined \def \showISBNx     #1{\unskip}     \fi
\ifx \showISBNxiii \undefined \def \showISBNxiii  #1{\unskip}     \fi
\ifx \showISSN     \undefined \def \showISSN      #1{\unskip}     \fi
\ifx \showLCCN     \undefined \def \showLCCN      #1{\unskip}     \fi
\ifx \shownote     \undefined \def \shownote      #1{#1}          \fi
\ifx \showarticletitle \undefined \def \showarticletitle #1{#1}   \fi
\ifx \showURL      \undefined \def \showURL       {\relax}        \fi
\providecommand\bibfield[2]{#2}
\providecommand\bibinfo[2]{#2}
\providecommand\natexlab[1]{#1}
\providecommand\showeprint[2][]{arXiv:#2}

\bibitem[Abdollahpouri et~al\mbox{.}(2017)]%
        {abdollahpouri2017controlling}
\bibfield{author}{\bibinfo{person}{Himan Abdollahpouri}, \bibinfo{person}{Robin
  Burke}, {and} \bibinfo{person}{Bamshad Mobasher}.}
  \bibinfo{year}{2017}\natexlab{}.
\newblock \showarticletitle{Controlling popularity bias in learning-to-rank
  recommendation}. In \bibinfo{booktitle}{\emph{Proceedings of the eleventh ACM
  conference on recommender systems}}. \bibinfo{pages}{42--46}.
\newblock


\bibitem[Begley et~al\mbox{.}(2020)]%
        {begley2020explainability}
\bibfield{author}{\bibinfo{person}{Tom Begley}, \bibinfo{person}{Tobias
  Schwedes}, \bibinfo{person}{Christopher Frye}, {and} \bibinfo{person}{Ilya
  Feige}.} \bibinfo{year}{2020}\natexlab{}.
\newblock \showarticletitle{Explainability for fair machine learning}.
\newblock \bibinfo{journal}{\emph{arXiv preprint arXiv:2010.07389}}
  (\bibinfo{year}{2020}).
\newblock


\bibitem[Beigi et~al\mbox{.}(2020)]%
        {beigi2020privacy}
\bibfield{author}{\bibinfo{person}{Ghazaleh Beigi}, \bibinfo{person}{Ahmadreza
  Mosallanezhad}, \bibinfo{person}{Ruocheng Guo}, \bibinfo{person}{Hamidreza
  Alvari}, \bibinfo{person}{Alexander Nou}, {and} \bibinfo{person}{Huan Liu}.}
  \bibinfo{year}{2020}\natexlab{}.
\newblock \showarticletitle{Privacy-aware recommendation with private-attribute
  protection using adversarial learning}. In
  \bibinfo{booktitle}{\emph{Proceedings of the 13th International Conference on
  Web Search and Data Mining}}. \bibinfo{pages}{34--42}.
\newblock


\bibitem[Beutel et~al\mbox{.}(2019)]%
        {beutel2019fairness}
\bibfield{author}{\bibinfo{person}{Alex Beutel}, \bibinfo{person}{Jilin Chen},
  \bibinfo{person}{Tulsee Doshi}, \bibinfo{person}{Hai Qian},
  \bibinfo{person}{Li Wei}, \bibinfo{person}{Yi Wu}, \bibinfo{person}{Lukasz
  Heldt}, \bibinfo{person}{Zhe Zhao}, \bibinfo{person}{Lichan Hong},
  \bibinfo{person}{Ed~H Chi}, {et~al\mbox{.}}} \bibinfo{year}{2019}\natexlab{}.
\newblock \showarticletitle{Fairness in recommendation ranking through pairwise
  comparisons}. In \bibinfo{booktitle}{\emph{Proceedings of the 25th ACM SIGKDD
  International Conference on Knowledge Discovery \& Data Mining}}.
  \bibinfo{pages}{2212--2220}.
\newblock


\bibitem[Bottou(2012)]%
        {bottou2012stochastic}
\bibfield{author}{\bibinfo{person}{L{\'e}on Bottou}.}
  \bibinfo{year}{2012}\natexlab{}.
\newblock \showarticletitle{Stochastic gradient descent tricks}.
\newblock In \bibinfo{booktitle}{\emph{Neural networks: Tricks of the trade}}.
  \bibinfo{publisher}{Springer}, \bibinfo{pages}{421--436}.
\newblock


\bibitem[Byrne(2007)]%
        {byrne2007rational}
\bibfield{author}{\bibinfo{person}{Ruth~MJ Byrne}.}
  \bibinfo{year}{2007}\natexlab{}.
\newblock \bibinfo{booktitle}{\emph{The rational imagination: How people create
  alternatives to reality}}.
\newblock \bibinfo{publisher}{MIT press}.
\newblock


\bibitem[Chen et~al\mbox{.}(2018b)]%
        {chen2018neural}
\bibfield{author}{\bibinfo{person}{Chong Chen}, \bibinfo{person}{Min Zhang},
  \bibinfo{person}{Yiqun Liu}, {and} \bibinfo{person}{Shaoping Ma}.}
  \bibinfo{year}{2018}\natexlab{b}.
\newblock \showarticletitle{Neural attentional rating regression with
  review-level explanations}. In \bibinfo{booktitle}{\emph{Proceedings of the
  2018 World Wide Web Conference}}. \bibinfo{pages}{1583--1592}.
\newblock


\bibitem[Chen et~al\mbox{.}(2018a)]%
        {chen2018investigating}
\bibfield{author}{\bibinfo{person}{Le Chen}, \bibinfo{person}{Ruijun Ma},
  \bibinfo{person}{Anik{\'o} Hann{\'a}k}, {and} \bibinfo{person}{Christo
  Wilson}.} \bibinfo{year}{2018}\natexlab{a}.
\newblock \showarticletitle{Investigating the impact of gender on rank in
  resume search engines}. In \bibinfo{booktitle}{\emph{Proceedings of the 2018
  chi conference on human factors in computing systems}}.
  \bibinfo{pages}{1--14}.
\newblock


\bibitem[Covington et~al\mbox{.}(2016)]%
        {covington2016deep}
\bibfield{author}{\bibinfo{person}{Paul Covington}, \bibinfo{person}{Jay
  Adams}, {and} \bibinfo{person}{Emre Sargin}.}
  \bibinfo{year}{2016}\natexlab{}.
\newblock \showarticletitle{Deep neural networks for youtube recommendations}.
  In \bibinfo{booktitle}{\emph{Proceedings of the 10th ACM conference on
  recommender systems}}. \bibinfo{pages}{191--198}.
\newblock


\bibitem[Dey and Salem(2017)]%
        {dey2017gate}
\bibfield{author}{\bibinfo{person}{Rahul Dey} {and} \bibinfo{person}{Fathi~M
  Salem}.} \bibinfo{year}{2017}\natexlab{}.
\newblock \showarticletitle{Gate-variants of gated recurrent unit (GRU) neural
  networks}. In \bibinfo{booktitle}{\emph{2017 IEEE 60th international midwest
  symposium on circuits and systems (MWSCAS)}}. IEEE,
  \bibinfo{pages}{1597--1600}.
\newblock


\bibitem[Diaz et~al\mbox{.}(2020)]%
        {diaz2020evaluating}
\bibfield{author}{\bibinfo{person}{Fernando Diaz}, \bibinfo{person}{Bhaskar
  Mitra}, \bibinfo{person}{Michael~D Ekstrand}, \bibinfo{person}{Asia~J Biega},
  {and} \bibinfo{person}{Ben Carterette}.} \bibinfo{year}{2020}\natexlab{}.
\newblock \showarticletitle{Evaluating stochastic rankings with expected
  exposure}. In \bibinfo{booktitle}{\emph{Proceedings of the 29th ACM
  international conference on information \& knowledge management}}.
  \bibinfo{pages}{275--284}.
\newblock


\bibitem[Fernandez et~al\mbox{.}(2020)]%
        {DBLP:journals/corr/abs-2001-07417}
\bibfield{author}{\bibinfo{person}{Carlos Fernandez},
  \bibinfo{person}{Foster~J. Provost}, {and} \bibinfo{person}{Xintian Han}.}
  \bibinfo{year}{2020}\natexlab{}.
\newblock \showarticletitle{Explaining Data-Driven Decisions made by {AI}
  Systems: The Counterfactual Approach}.
\newblock \bibinfo{journal}{\emph{CoRR}}  \bibinfo{volume}{abs/2001.07417}
  (\bibinfo{year}{2020}).
\newblock
\showeprint[arXiv]{2001.07417}
\urldef\tempurl%
\url{https://arxiv.org/abs/2001.07417}
\showURL{%
\tempurl}


\bibitem[Fu et~al\mbox{.}(2020)]%
        {fu2020fairness}
\bibfield{author}{\bibinfo{person}{Zuohui Fu}, \bibinfo{person}{Yikun Xian},
  \bibinfo{person}{Ruoyuan Gao}, \bibinfo{person}{Jieyu Zhao},
  \bibinfo{person}{Qiaoying Huang}, \bibinfo{person}{Yingqiang Ge},
  \bibinfo{person}{Shuyuan Xu}, \bibinfo{person}{Shijie Geng},
  \bibinfo{person}{Chirag Shah}, \bibinfo{person}{Yongfeng Zhang},
  {et~al\mbox{.}}} \bibinfo{year}{2020}\natexlab{}.
\newblock \showarticletitle{Fairness-aware explainable recommendation over
  knowledge graphs}. In \bibinfo{booktitle}{\emph{Proceedings of the 43rd
  International ACM SIGIR Conference on Research and Development in Information
  Retrieval}}. \bibinfo{pages}{69--78}.
\newblock


\bibitem[Ge et~al\mbox{.}(2021a)]%
        {ge2021towards}
\bibfield{author}{\bibinfo{person}{Yingqiang Ge}, \bibinfo{person}{Shuchang
  Liu}, \bibinfo{person}{Ruoyuan Gao}, \bibinfo{person}{Yikun Xian},
  \bibinfo{person}{Yunqi Li}, \bibinfo{person}{Xiangyu Zhao},
  \bibinfo{person}{Changhua Pei}, \bibinfo{person}{Fei Sun},
  \bibinfo{person}{Junfeng Ge}, \bibinfo{person}{Wenwu Ou}, {et~al\mbox{.}}}
  \bibinfo{year}{2021}\natexlab{a}.
\newblock \showarticletitle{Towards long-term fairness in recommendation}. In
  \bibinfo{booktitle}{\emph{Proceedings of the 14th ACM International
  Conference on Web Search and Data Mining}}. \bibinfo{pages}{445--453}.
\newblock


\bibitem[Ge et~al\mbox{.}(2021b)]%
        {ge2021counterfactual}
\bibfield{author}{\bibinfo{person}{Yingqiang Ge}, \bibinfo{person}{Shuchang
  Liu}, \bibinfo{person}{Zelong Li}, \bibinfo{person}{Shuyuan Xu},
  \bibinfo{person}{Shijie Geng}, \bibinfo{person}{Yunqi Li},
  \bibinfo{person}{Juntao Tan}, \bibinfo{person}{Fei Sun}, {and}
  \bibinfo{person}{Yongfeng Zhang}.} \bibinfo{year}{2021}\natexlab{b}.
\newblock \showarticletitle{Counterfactual Evaluation for Explainable AI}.
\newblock \bibinfo{journal}{\emph{arXiv preprint arXiv:2109.01962}}
  (\bibinfo{year}{2021}).
\newblock


\bibitem[Ge et~al\mbox{.}(2022a)]%
        {DBLP:conf/sigir/GeTZXL0FGLZ22}
\bibfield{author}{\bibinfo{person}{Yingqiang Ge}, \bibinfo{person}{Juntao Tan},
  \bibinfo{person}{Yan Zhu}, \bibinfo{person}{Yinglong Xia},
  \bibinfo{person}{Jiebo Luo}, \bibinfo{person}{Shuchang Liu},
  \bibinfo{person}{Zuohui Fu}, \bibinfo{person}{Shijie Geng},
  \bibinfo{person}{Zelong Li}, {and} \bibinfo{person}{Yongfeng Zhang}.}
  \bibinfo{year}{2022}\natexlab{a}.
\newblock \showarticletitle{Explainable Fairness in Recommendation}. In
  \bibinfo{booktitle}{\emph{{SIGIR} '22: The 45th International {ACM} {SIGIR}
  Conference on Research and Development in Information Retrieval, Madrid,
  Spain, July 11 - 15, 2022}}, \bibfield{editor}{\bibinfo{person}{Enrique
  Amig{\'{o}}}, \bibinfo{person}{Pablo Castells}, \bibinfo{person}{Julio
  Gonzalo}, \bibinfo{person}{Ben Carterette}, \bibinfo{person}{J.~Shane
  Culpepper}, {and} \bibinfo{person}{Gabriella Kazai}} (Eds.).
  \bibinfo{publisher}{{ACM}}, \bibinfo{pages}{681--691}.
\newblock
\urldef\tempurl%
\url{https://doi.org/10.1145/3477495.3531973}
\showDOI{\tempurl}


\bibitem[Ge et~al\mbox{.}(2022b)]%
        {ge2022toward}
\bibfield{author}{\bibinfo{person}{Yingqiang Ge}, \bibinfo{person}{Xiaoting
  Zhao}, \bibinfo{person}{Lucia Yu}, \bibinfo{person}{Saurabh Paul},
  \bibinfo{person}{Diane Hu}, \bibinfo{person}{Chu-Cheng Hsieh}, {and}
  \bibinfo{person}{Yongfeng Zhang}.} \bibinfo{year}{2022}\natexlab{b}.
\newblock \showarticletitle{Toward Pareto Efficient Fairness-Utility Trade-off
  in Recommendation through Reinforcement Learning}. In
  \bibinfo{booktitle}{\emph{Proceedings of the Fifteenth ACM International
  Conference on Web Search and Data Mining}}. \bibinfo{pages}{316--324}.
\newblock


\bibitem[Ghazimatin et~al\mbox{.}(2020)]%
        {ghazimatin2020prince}
\bibfield{author}{\bibinfo{person}{Azin Ghazimatin}, \bibinfo{person}{Oana
  Balalau}, \bibinfo{person}{Rishiraj Saha~Roy}, {and} \bibinfo{person}{Gerhard
  Weikum}.} \bibinfo{year}{2020}\natexlab{}.
\newblock \showarticletitle{PRINCE: Provider-side interpretability with
  counterfactual explanations in recommender systems}. In
  \bibinfo{booktitle}{\emph{Proceedings of the 13th International Conference on
  Web Search and Data Mining}}. \bibinfo{pages}{196--204}.
\newblock


\bibitem[Goyani and Chaurasiya(2020)]%
        {goyani2020review}
\bibfield{author}{\bibinfo{person}{Mahesh Goyani} {and} \bibinfo{person}{Neha
  Chaurasiya}.} \bibinfo{year}{2020}\natexlab{}.
\newblock \showarticletitle{A review of movie recommendation system:
  Limitations, Survey and Challenges}.
\newblock \bibinfo{journal}{\emph{ELCVIA: electronic letters on computer vision
  and image analysis}} \bibinfo{volume}{19}, \bibinfo{number}{3}
  (\bibinfo{year}{2020}), \bibinfo{pages}{0018--37}.
\newblock


\bibitem[Guan et~al\mbox{.}(2010)]%
        {guan2010document}
\bibfield{author}{\bibinfo{person}{Ziyu Guan}, \bibinfo{person}{Can Wang},
  \bibinfo{person}{Jiajun Bu}, \bibinfo{person}{Chun Chen},
  \bibinfo{person}{Kun Yang}, \bibinfo{person}{Deng Cai}, {and}
  \bibinfo{person}{Xiaofei He}.} \bibinfo{year}{2010}\natexlab{}.
\newblock \showarticletitle{Document recommendation in social tagging
  services}. In \bibinfo{booktitle}{\emph{Proceedings of the 19th international
  conference on World wide web}}. \bibinfo{pages}{391--400}.
\newblock


\bibitem[Hamilton et~al\mbox{.}(2017)]%
        {hamilton2017inductive}
\bibfield{author}{\bibinfo{person}{Will Hamilton}, \bibinfo{person}{Zhitao
  Ying}, {and} \bibinfo{person}{Jure Leskovec}.}
  \bibinfo{year}{2017}\natexlab{}.
\newblock \showarticletitle{Inductive representation learning on large graphs}.
\newblock \bibinfo{journal}{\emph{Advances in neural information processing
  systems}}  \bibinfo{volume}{30} (\bibinfo{year}{2017}).
\newblock


\bibitem[He et~al\mbox{.}(2021)]%
        {he2021click}
\bibfield{author}{\bibinfo{person}{Li He}, \bibinfo{person}{Hongxu Chen},
  \bibinfo{person}{Dingxian Wang}, \bibinfo{person}{Shoaib Jameel},
  \bibinfo{person}{Philip Yu}, {and} \bibinfo{person}{Guandong Xu}.}
  \bibinfo{year}{2021}\natexlab{}.
\newblock \showarticletitle{Click-Through Rate Prediction with Multi-Modal
  Hypergraphs}. In \bibinfo{booktitle}{\emph{Proceedings of the 30th ACM
  International Conference on Information \& Knowledge Management}}.
  \bibinfo{pages}{690--699}.
\newblock


\bibitem[He et~al\mbox{.}(2017)]%
        {he2017neural}
\bibfield{author}{\bibinfo{person}{Xiangnan He}, \bibinfo{person}{Lizi Liao},
  \bibinfo{person}{Hanwang Zhang}, \bibinfo{person}{Liqiang Nie},
  \bibinfo{person}{Xia Hu}, {and} \bibinfo{person}{Tat-Seng Chua}.}
  \bibinfo{year}{2017}\natexlab{}.
\newblock \showarticletitle{Neural collaborative filtering}. In
  \bibinfo{booktitle}{\emph{Proceedings of the 26th international conference on
  world wide web}}. \bibinfo{pages}{173--182}.
\newblock


\bibitem[Hu et~al\mbox{.}(2018)]%
        {hu2018leveraging}
\bibfield{author}{\bibinfo{person}{Binbin Hu}, \bibinfo{person}{Chuan Shi},
  \bibinfo{person}{Wayne~Xin Zhao}, {and} \bibinfo{person}{Philip~S Yu}.}
  \bibinfo{year}{2018}\natexlab{}.
\newblock \showarticletitle{Leveraging meta-path based context for top-n
  recommendation with a neural co-attention model}. In
  \bibinfo{booktitle}{\emph{Proceedings of the 24th ACM SIGKDD International
  Conference on Knowledge Discovery \& Data Mining}}.
  \bibinfo{pages}{1531--1540}.
\newblock


\bibitem[Li et~al\mbox{.}(2022c)]%
        {10.1145/3533725}
\bibfield{author}{\bibinfo{person}{Qian Li}, \bibinfo{person}{Xiangmeng Wang},
  \bibinfo{person}{Zhichao Wang}, {and} \bibinfo{person}{Guandong Xu}.}
  \bibinfo{year}{2022}\natexlab{c}.
\newblock \showarticletitle{Be Causal: De-Biasing Social Network Confounding in
  Recommendation}.
\newblock \bibinfo{journal}{\emph{ACM Trans. Knowl. Discov. Data}}
  (\bibinfo{date}{apr} \bibinfo{year}{2022}).
\newblock
\showISSN{1556-4681}
\urldef\tempurl%
\url{https://doi.org/10.1145/3533725}
\showDOI{\tempurl}


\bibitem[Li et~al\mbox{.}(2022b)]%
        {li2022deep}
\bibfield{author}{\bibinfo{person}{Qian Li}, \bibinfo{person}{Zhichao Wang},
  \bibinfo{person}{Shaowu Liu}, \bibinfo{person}{Gang Li}, {and}
  \bibinfo{person}{Guandong Xu}.} \bibinfo{year}{2022}\natexlab{b}.
\newblock \showarticletitle{Deep treatment-adaptive network for causal
  inference}.
\newblock \bibinfo{journal}{\emph{The VLDB Journal}} (\bibinfo{year}{2022}),
  \bibinfo{pages}{1--16}.
\newblock


\bibitem[Li et~al\mbox{.}(2021a)]%
        {li2021user}
\bibfield{author}{\bibinfo{person}{Yunqi Li}, \bibinfo{person}{Hanxiong Chen},
  \bibinfo{person}{Zuohui Fu}, \bibinfo{person}{Yingqiang Ge}, {and}
  \bibinfo{person}{Yongfeng Zhang}.} \bibinfo{year}{2021}\natexlab{a}.
\newblock \showarticletitle{User-oriented fairness in recommendation}. In
  \bibinfo{booktitle}{\emph{Proceedings of the Web Conference 2021}}.
  \bibinfo{pages}{624--632}.
\newblock


\bibitem[Li et~al\mbox{.}(2022a)]%
        {li2022fairness}
\bibfield{author}{\bibinfo{person}{Yunqi Li}, \bibinfo{person}{Hanxiong Chen},
  \bibinfo{person}{Shuyuan Xu}, \bibinfo{person}{Yingqiang Ge},
  \bibinfo{person}{Juntao Tan}, \bibinfo{person}{Shuchang Liu}, {and}
  \bibinfo{person}{Yongfeng Zhang}.} \bibinfo{year}{2022}\natexlab{a}.
\newblock \showarticletitle{Fairness in Recommendation: A Survey}.
\newblock \bibinfo{journal}{\emph{arXiv preprint arXiv:2205.13619}}
  (\bibinfo{year}{2022}).
\newblock


\bibitem[Li et~al\mbox{.}(2021b)]%
        {li2021towards}
\bibfield{author}{\bibinfo{person}{Yunqi Li}, \bibinfo{person}{Hanxiong Chen},
  \bibinfo{person}{Shuyuan Xu}, \bibinfo{person}{Yingqiang Ge}, {and}
  \bibinfo{person}{Yongfeng Zhang}.} \bibinfo{year}{2021}\natexlab{b}.
\newblock \showarticletitle{Towards personalized fairness based on causal
  notion}. In \bibinfo{booktitle}{\emph{Proceedings of the 44th International
  ACM SIGIR Conference on Research and Development in Information Retrieval}}.
  \bibinfo{pages}{1054--1063}.
\newblock


\bibitem[Liu et~al\mbox{.}(2020)]%
        {liu2020balancing}
\bibfield{author}{\bibinfo{person}{Weiwen Liu}, \bibinfo{person}{Feng Liu},
  \bibinfo{person}{Ruiming Tang}, \bibinfo{person}{Ben Liao},
  \bibinfo{person}{Guangyong Chen}, {and} \bibinfo{person}{Pheng~Ann Heng}.}
  \bibinfo{year}{2020}\natexlab{}.
\newblock \showarticletitle{Balancing between accuracy and fairness for
  interactive recommendation with reinforcement learning}. In
  \bibinfo{booktitle}{\emph{Pacific-asia conference on knowledge discovery and
  data mining}}. Springer, \bibinfo{pages}{155--167}.
\newblock


\bibitem[Nguyen and Bai(2010)]%
        {nguyen2010cosine}
\bibfield{author}{\bibinfo{person}{Hieu~V Nguyen} {and} \bibinfo{person}{Li
  Bai}.} \bibinfo{year}{2010}\natexlab{}.
\newblock \showarticletitle{Cosine similarity metric learning for face
  verification}. In \bibinfo{booktitle}{\emph{Asian conference on computer
  vision}}. Springer, \bibinfo{pages}{709--720}.
\newblock


\bibitem[Nguyen et~al\mbox{.}(2014)]%
        {nguyen2014exploring}
\bibfield{author}{\bibinfo{person}{Tien~T Nguyen}, \bibinfo{person}{Pik-Mai
  Hui}, \bibinfo{person}{F~Maxwell Harper}, \bibinfo{person}{Loren Terveen},
  {and} \bibinfo{person}{Joseph~A Konstan}.} \bibinfo{year}{2014}\natexlab{}.
\newblock \showarticletitle{Exploring the filter bubble: the effect of using
  recommender systems on content diversity}. In
  \bibinfo{booktitle}{\emph{Proceedings of the 23rd international conference on
  World wide web}}. \bibinfo{pages}{677--686}.
\newblock


\bibitem[Perc(2014)]%
        {perc2014matthew}
\bibfield{author}{\bibinfo{person}{Matja{\v{z}} Perc}.}
  \bibinfo{year}{2014}\natexlab{}.
\newblock \showarticletitle{The Matthew effect in empirical data}.
\newblock \bibinfo{journal}{\emph{Journal of The Royal Society Interface}}
  \bibinfo{volume}{11}, \bibinfo{number}{98} (\bibinfo{year}{2014}),
  \bibinfo{pages}{20140378}.
\newblock


\bibitem[Reddy et~al\mbox{.}(2019)]%
        {reddy2019content}
\bibfield{author}{\bibinfo{person}{SRS Reddy}, \bibinfo{person}{Sravani
  Nalluri}, \bibinfo{person}{Subramanyam Kunisetti}, \bibinfo{person}{S Ashok},
  {and} \bibinfo{person}{B Venkatesh}.} \bibinfo{year}{2019}\natexlab{}.
\newblock \showarticletitle{Content-based movie recommendation system using
  genre correlation}. In \bibinfo{booktitle}{\emph{Smart Intelligent Computing
  and Applications: Proceedings of the Second International Conference on SCI
  2018, Volume 2}}. Springer, \bibinfo{pages}{391--397}.
\newblock


\bibitem[Rendle et~al\mbox{.}(2012)]%
        {rendle2012bpr}
\bibfield{author}{\bibinfo{person}{Steffen Rendle}, \bibinfo{person}{Christoph
  Freudenthaler}, \bibinfo{person}{Zeno Gantner}, {and} \bibinfo{person}{Lars
  Schmidt-Thieme}.} \bibinfo{year}{2012}\natexlab{}.
\newblock \showarticletitle{BPR: Bayesian personalized ranking from implicit
  feedback}.
\newblock \bibinfo{journal}{\emph{arXiv preprint arXiv:1205.2618}}
  (\bibinfo{year}{2012}).
\newblock


\bibitem[Resheff et~al\mbox{.}(2018)]%
        {resheff2018privacy}
\bibfield{author}{\bibinfo{person}{Yehezkel~S Resheff}, \bibinfo{person}{Yanai
  Elazar}, \bibinfo{person}{Moni Shahar}, {and} \bibinfo{person}{Oren~Sar
  Shalom}.} \bibinfo{year}{2018}\natexlab{}.
\newblock \showarticletitle{Privacy and fairness in recommender systems via
  adversarial training of user representations}.
\newblock \bibinfo{journal}{\emph{arXiv preprint arXiv:1807.03521}}
  (\bibinfo{year}{2018}).
\newblock


\bibitem[Shi et~al\mbox{.}(2018)]%
        {shi2018heterogeneous}
\bibfield{author}{\bibinfo{person}{Chuan Shi}, \bibinfo{person}{Binbin Hu},
  \bibinfo{person}{Wayne~Xin Zhao}, {and} \bibinfo{person}{S~Yu Philip}.}
  \bibinfo{year}{2018}\natexlab{}.
\newblock \showarticletitle{Heterogeneous information network embedding for
  recommendation}.
\newblock \bibinfo{journal}{\emph{IEEE Transactions on Knowledge and Data
  Engineering}} \bibinfo{volume}{31}, \bibinfo{number}{2}
  (\bibinfo{year}{2018}), \bibinfo{pages}{357--370}.
\newblock


\bibitem[Shi et~al\mbox{.}(2019)]%
        {shi2019semrec}
\bibfield{author}{\bibinfo{person}{Chuan Shi}, \bibinfo{person}{Zhiqiang
  Zhang}, \bibinfo{person}{Yugang Ji}, \bibinfo{person}{Weipeng Wang},
  \bibinfo{person}{Philip~S Yu}, {and} \bibinfo{person}{Zhiping Shi}.}
  \bibinfo{year}{2019}\natexlab{}.
\newblock \showarticletitle{SemRec: a personalized semantic recommendation
  method based on weighted heterogeneous information networks}.
\newblock \bibinfo{journal}{\emph{World Wide Web}}  \bibinfo{volume}{22}
  (\bibinfo{year}{2019}), \bibinfo{pages}{153--184}.
\newblock


\bibitem[Sutton et~al\mbox{.}(1999)]%
        {sutton1999policy}
\bibfield{author}{\bibinfo{person}{Richard~S Sutton}, \bibinfo{person}{David
  McAllester}, \bibinfo{person}{Satinder Singh}, {and} \bibinfo{person}{Yishay
  Mansour}.} \bibinfo{year}{1999}\natexlab{}.
\newblock \showarticletitle{Policy gradient methods for reinforcement learning
  with function approximation}.
\newblock \bibinfo{journal}{\emph{Advances in neural information processing
  systems}}  \bibinfo{volume}{12} (\bibinfo{year}{1999}).
\newblock


\bibitem[Swaminathan and Joachims(2015)]%
        {swaminathan2015counterfactual}
\bibfield{author}{\bibinfo{person}{Adith Swaminathan} {and}
  \bibinfo{person}{Thorsten Joachims}.} \bibinfo{year}{2015}\natexlab{}.
\newblock \showarticletitle{Counterfactual risk minimization: Learning from
  logged bandit feedback}. In \bibinfo{booktitle}{\emph{International
  Conference on Machine Learning}}. PMLR, \bibinfo{pages}{814--823}.
\newblock


\bibitem[Tran et~al\mbox{.}(2021)]%
        {tran2021counterfactual}
\bibfield{author}{\bibinfo{person}{Khanh~Hiep Tran}, \bibinfo{person}{Azin
  Ghazimatin}, {and} \bibinfo{person}{Rishiraj Saha~Roy}.}
  \bibinfo{year}{2021}\natexlab{}.
\newblock \showarticletitle{Counterfactual Explanations for Neural
  Recommenders}. In \bibinfo{booktitle}{\emph{Proceedings of the 44th
  International ACM SIGIR Conference on Research and Development in Information
  Retrieval}}. \bibinfo{pages}{1627--1631}.
\newblock


\bibitem[Vaswani et~al\mbox{.}(2017)]%
        {vaswani2017attention}
\bibfield{author}{\bibinfo{person}{Ashish Vaswani}, \bibinfo{person}{Noam
  Shazeer}, \bibinfo{person}{Niki Parmar}, \bibinfo{person}{Jakob Uszkoreit},
  \bibinfo{person}{Llion Jones}, \bibinfo{person}{Aidan~N Gomez},
  \bibinfo{person}{{\L}ukasz Kaiser}, {and} \bibinfo{person}{Illia
  Polosukhin}.} \bibinfo{year}{2017}\natexlab{}.
\newblock \showarticletitle{Attention is all you need}.
\newblock \bibinfo{journal}{\emph{Advances in neural information processing
  systems}}  \bibinfo{volume}{30} (\bibinfo{year}{2017}).
\newblock


\bibitem[Verma et~al\mbox{.}(2020)]%
        {verma2020counterfactual}
\bibfield{author}{\bibinfo{person}{Sahil Verma}, \bibinfo{person}{John
  Dickerson}, {and} \bibinfo{person}{Keegan Hines}.}
  \bibinfo{year}{2020}\natexlab{}.
\newblock \showarticletitle{Counterfactual explanations for machine learning: A
  review}.
\newblock \bibinfo{journal}{\emph{arXiv preprint arXiv:2010.10596}}
  (\bibinfo{year}{2020}).
\newblock


\bibitem[Wang et~al\mbox{.}(2019a)]%
        {wang2019kgat}
\bibfield{author}{\bibinfo{person}{Xiang Wang}, \bibinfo{person}{Xiangnan He},
  \bibinfo{person}{Yixin Cao}, \bibinfo{person}{Meng Liu}, {and}
  \bibinfo{person}{Tat-Seng Chua}.} \bibinfo{year}{2019}\natexlab{a}.
\newblock \showarticletitle{Kgat: Knowledge graph attention network for
  recommendation}. In \bibinfo{booktitle}{\emph{Proceedings of the 25th ACM
  SIGKDD international conference on knowledge discovery \& data mining}}.
  \bibinfo{pages}{950--958}.
\newblock


\bibitem[Wang et~al\mbox{.}(2019b)]%
        {wang2019heterogeneous}
\bibfield{author}{\bibinfo{person}{Xiao Wang}, \bibinfo{person}{Houye Ji},
  \bibinfo{person}{Chuan Shi}, \bibinfo{person}{Bai Wang},
  \bibinfo{person}{Yanfang Ye}, \bibinfo{person}{Peng Cui}, {and}
  \bibinfo{person}{Philip~S Yu}.} \bibinfo{year}{2019}\natexlab{b}.
\newblock \showarticletitle{Heterogeneous graph attention network}. In
  \bibinfo{booktitle}{\emph{The World Wide Web Conference}}.
  \bibinfo{pages}{2022--2032}.
\newblock


\bibitem[Wang et~al\mbox{.}(2022c)]%
        {wang2022causal}
\bibfield{author}{\bibinfo{person}{Xiangmeng Wang}, \bibinfo{person}{Qian Li},
  \bibinfo{person}{Dianer Yu}, \bibinfo{person}{Peng Cui},
  \bibinfo{person}{Zhichao Wang}, {and} \bibinfo{person}{Guandong Xu}.}
  \bibinfo{year}{2022}\natexlab{c}.
\newblock \showarticletitle{Causal Disentanglement for Semantics-Aware Intent
  Learning in Recommendation}.
\newblock \bibinfo{journal}{\emph{IEEE Transactions on Knowledge and Data
  Engineering}} (\bibinfo{year}{2022}).
\newblock


\bibitem[Wang et~al\mbox{.}(2022d)]%
        {10.1145/3477495.3532021}
\bibfield{author}{\bibinfo{person}{Xiangmeng Wang}, \bibinfo{person}{Qian Li},
  \bibinfo{person}{Dianer Yu}, \bibinfo{person}{Zhichao Wang},
  \bibinfo{person}{Hongxu Chen}, {and} \bibinfo{person}{Guandong Xu}.}
  \bibinfo{year}{2022}\natexlab{d}.
\newblock \showarticletitle{MGPolicy: Meta Graph Enhanced Off-Policy Learning
  for Recommendations}. In \bibinfo{booktitle}{\emph{Proceedings of the 45th
  International ACM SIGIR Conference on Research and Development in Information
  Retrieval}} (Madrid, Spain) \emph{(\bibinfo{series}{SIGIR '22})}.
  \bibinfo{publisher}{Association for Computing Machinery},
  \bibinfo{address}{New York, NY, USA}, \bibinfo{pages}{1369–1378}.
\newblock
\showISBNx{9781450387323}
\urldef\tempurl%
\url{https://doi.org/10.1145/3477495.3532021}
\showDOI{\tempurl}


\bibitem[Wang et~al\mbox{.}(2022a)]%
        {10.1145/3485447.3512072}
\bibfield{author}{\bibinfo{person}{Xiangmeng Wang}, \bibinfo{person}{Qian Li},
  \bibinfo{person}{Dianer Yu}, {and} \bibinfo{person}{Guandong Xu}.}
  \bibinfo{year}{2022}\natexlab{a}.
\newblock \showarticletitle{Off-Policy Learning over Heterogeneous Information
  for Recommendation} \emph{(\bibinfo{series}{WWW '22})}.
  \bibinfo{publisher}{Association for Computing Machinery},
  \bibinfo{address}{New York, NY, USA}, \bibinfo{pages}{2348–2359}.
\newblock
\showISBNx{9781450390965}
\urldef\tempurl%
\url{https://doi.org/10.1145/3485447.3512072}
\showDOI{\tempurl}


\bibitem[Wang et~al\mbox{.}(2022b)]%
        {wang2022reinforced}
\bibfield{author}{\bibinfo{person}{Xiangmeng Wang}, \bibinfo{person}{Qian Li},
  \bibinfo{person}{Dianer Yu}, {and} \bibinfo{person}{Guandong Xu}.}
  \bibinfo{year}{2022}\natexlab{b}.
\newblock \showarticletitle{Reinforced Path Reasoning for Counterfactual
  Explainable Recommendation}.
\newblock \bibinfo{journal}{\emph{arXiv preprint arXiv:2207.06674}}
  (\bibinfo{year}{2022}).
\newblock


\bibitem[Wang et~al\mbox{.}(2020)]%
        {wang2020joint}
\bibfield{author}{\bibinfo{person}{Xiangmeng Wang}, \bibinfo{person}{Qian Li},
  \bibinfo{person}{Wu Zhang}, \bibinfo{person}{Guandong Xu},
  \bibinfo{person}{Shaowu Liu}, {and} \bibinfo{person}{Wenhao Zhu}.}
  \bibinfo{year}{2020}\natexlab{}.
\newblock \showarticletitle{Joint relational dependency learning for sequential
  recommendation}. In \bibinfo{booktitle}{\emph{Pacific-Asia Conference on
  Knowledge Discovery and Data Mining}}. Springer, \bibinfo{pages}{168--180}.
\newblock


\bibitem[Wang et~al\mbox{.}(2022e)]%
        {wang2022survey2}
\bibfield{author}{\bibinfo{person}{Yifan Wang}, \bibinfo{person}{Weizhi Ma},
  \bibinfo{person}{Min Zhang*}, \bibinfo{person}{Yiqun Liu}, {and}
  \bibinfo{person}{Shaoping Ma}.} \bibinfo{year}{2022}\natexlab{e}.
\newblock \showarticletitle{A Survey on the Fairness of Recommender Systems}.
\newblock \bibinfo{journal}{\emph{ACM Journal of the ACM (JACM)}}
  (\bibinfo{year}{2022}).
\newblock


\bibitem[Williams(1992)]%
        {williams1992simple}
\bibfield{author}{\bibinfo{person}{Ronald~J Williams}.}
  \bibinfo{year}{1992}\natexlab{}.
\newblock \showarticletitle{Simple statistical gradient-following algorithms
  for connectionist reinforcement learning}.
\newblock \bibinfo{journal}{\emph{Machine learning}} \bibinfo{volume}{8},
  \bibinfo{number}{3} (\bibinfo{year}{1992}), \bibinfo{pages}{229--256}.
\newblock


\bibitem[Williamson and Forbes(2014)]%
        {williamson2014introduction}
\bibfield{author}{\bibinfo{person}{Elizabeth~J Williamson} {and}
  \bibinfo{person}{Andrew Forbes}.} \bibinfo{year}{2014}\natexlab{}.
\newblock \showarticletitle{Introduction to propensity scores}.
\newblock \bibinfo{journal}{\emph{Respirology}} \bibinfo{volume}{19},
  \bibinfo{number}{5} (\bibinfo{year}{2014}), \bibinfo{pages}{625--635}.
\newblock


\bibitem[Woodward(2004)]%
        {woodwardMakingThingsHappen2004}
\bibfield{author}{\bibinfo{person}{James Woodward}.}
  \bibinfo{year}{2004}\natexlab{}.
\newblock \bibinfo{booktitle}{\emph{Making {Things} {Happen}: {A} {Theory} of
  {Causal} {Explanation}}}.
\newblock \bibinfo{publisher}{Oxford University Press}, \bibinfo{address}{New
  York}.
\newblock
\showISBNx{978-0-19-515527-3}
\urldef\tempurl%
\url{https://doi.org/10.1093/0195155270.001.0001}
\showDOI{\tempurl}


\bibitem[Xiong et~al\mbox{.}(2021)]%
        {xiong2021counterfactual}
\bibfield{author}{\bibinfo{person}{Kun Xiong}, \bibinfo{person}{Wenwen Ye},
  \bibinfo{person}{Xu Chen}, \bibinfo{person}{Yongfeng Zhang},
  \bibinfo{person}{Wayne~Xin Zhao}, \bibinfo{person}{Binbin Hu},
  \bibinfo{person}{Zhiqiang Zhang}, {and} \bibinfo{person}{Jun Zhou}.}
  \bibinfo{year}{2021}\natexlab{}.
\newblock \showarticletitle{Counterfactual Review-based Recommendation}. In
  \bibinfo{booktitle}{\emph{Proceedings of the 30th ACM International
  Conference on Information \& Knowledge Management}}.
  \bibinfo{pages}{2231--2240}.
\newblock


\bibitem[Xu et~al\mbox{.}(2020)]%
        {xu2020reluplex}
\bibfield{author}{\bibinfo{person}{Jin Xu}, \bibinfo{person}{Zishan Li},
  \bibinfo{person}{Bowen Du}, \bibinfo{person}{Miaomiao Zhang}, {and}
  \bibinfo{person}{Jing Liu}.} \bibinfo{year}{2020}\natexlab{}.
\newblock \showarticletitle{Reluplex made more practical: Leaky ReLU}. In
  \bibinfo{booktitle}{\emph{2020 IEEE Symposium on Computers and communications
  (ISCC)}}. IEEE, \bibinfo{pages}{1--7}.
\newblock


\bibitem[Xu et~al\mbox{.}(2018)]%
        {xu2018representation}
\bibfield{author}{\bibinfo{person}{Keyulu Xu}, \bibinfo{person}{Chengtao Li},
  \bibinfo{person}{Yonglong Tian}, \bibinfo{person}{Tomohiro Sonobe},
  \bibinfo{person}{Ken-ichi Kawarabayashi}, {and} \bibinfo{person}{Stefanie
  Jegelka}.} \bibinfo{year}{2018}\natexlab{}.
\newblock \showarticletitle{Representation learning on graphs with jumping
  knowledge networks}. In \bibinfo{booktitle}{\emph{International Conference on
  Machine Learning}}. PMLR, \bibinfo{pages}{5453--5462}.
\newblock


\bibitem[Yao and Huang(2017)]%
        {yao2017beyond}
\bibfield{author}{\bibinfo{person}{Sirui Yao} {and} \bibinfo{person}{Bert
  Huang}.} \bibinfo{year}{2017}\natexlab{}.
\newblock \showarticletitle{Beyond parity: Fairness objectives for
  collaborative filtering}.
\newblock \bibinfo{journal}{\emph{Advances in neural information processing
  systems}}  \bibinfo{volume}{30} (\bibinfo{year}{2017}).
\newblock


\bibitem[Yu et~al\mbox{.}(2022)]%
        {yu2022semantics}
\bibfield{author}{\bibinfo{person}{Dianer Yu}, \bibinfo{person}{Qian Li},
  \bibinfo{person}{Xiangmeng Wang}, \bibinfo{person}{Zhichao Wang},
  \bibinfo{person}{Yanan Cao}, {and} \bibinfo{person}{Guandong Xu}.}
  \bibinfo{year}{2022}\natexlab{}.
\newblock \showarticletitle{Semantics-Guided Disentangled Learning for
  Recommendation}. In \bibinfo{booktitle}{\emph{Pacific-Asia Conference on
  Knowledge Discovery and Data Mining}}. Springer, \bibinfo{pages}{249--261}.
\newblock


\bibitem[Zhang et~al\mbox{.}(2019)]%
        {zhang2019stylistic}
\bibfield{author}{\bibinfo{person}{Suiyun Zhang}, \bibinfo{person}{Zhizhong
  Han}, \bibinfo{person}{Yu-Kun Lai}, \bibinfo{person}{Matthias Zwicker}, {and}
  \bibinfo{person}{Hui Zhang}.} \bibinfo{year}{2019}\natexlab{}.
\newblock \showarticletitle{Stylistic scene enhancement GAN: mixed stylistic
  enhancement generation for 3D indoor scenes}.
\newblock \bibinfo{journal}{\emph{The Visual Computer}} \bibinfo{volume}{35},
  \bibinfo{number}{6} (\bibinfo{year}{2019}), \bibinfo{pages}{1157--1169}.
\newblock


\bibitem[Zhang and Sabuncu(2018)]%
        {zhang2018generalized}
\bibfield{author}{\bibinfo{person}{Zhilu Zhang} {and} \bibinfo{person}{Mert
  Sabuncu}.} \bibinfo{year}{2018}\natexlab{}.
\newblock \showarticletitle{Generalized cross entropy loss for training deep
  neural networks with noisy labels}.
\newblock \bibinfo{journal}{\emph{Advances in neural information processing
  systems}}  \bibinfo{volume}{31} (\bibinfo{year}{2018}).
\newblock


\bibitem[Zhao et~al\mbox{.}(2016)]%
        {zhao2016matrix}
\bibfield{author}{\bibinfo{person}{Lili Zhao}, \bibinfo{person}{Zhongqi Lu},
  \bibinfo{person}{Sinno~Jialin Pan}, \bibinfo{person}{Qiang Yang}, {and}
  \bibinfo{person}{Wei Xu}.} \bibinfo{year}{2016}\natexlab{}.
\newblock \showarticletitle{Matrix factorization+ for movie recommendation.}.
  In \bibinfo{booktitle}{\emph{IJCAI}}. \bibinfo{pages}{3945--3951}.
\newblock


\end{thebibliography}

\end{document}